\documentclass[11pt]{article}
\usepackage{amssymb}
\usepackage{amsmath}
\usepackage{graphicx,color}
\usepackage{wrapfig,framed}

\newcommand{\diag}{{\rm diag}\hskip 0.5truemm}

\newtheorem{theorem}{Theorem}

\newtheorem{prop}[theorem]{Proposition}

\newtheorem{lemma}[theorem]{Lemma}

\newtheorem{cor}[theorem]{Corollary}

\newtheorem{remark}[theorem]{Remark}
\newtheorem{defi}[theorem]{Definition}
\newtheorem{primer}[theorem]{Example}

\newcommand{\eqa}{\begin{eqnarray}}
\newcommand{\eeqa}{\end{eqnarray}}
\newcommand{\beq}{\begin{equation}}
\newcommand{\eeq}{\end{equation}}
\newcommand{\nn}{\nonumber}

\newcommand{\pal}{\partial}

\newcommand{\rea}{{\rm Re}\hskip 0.5truemm}
\newcommand{\ima}{{\rm Im}\hskip 0.5truemm}

\newcommand{\tr}{{\rm tr}}

\newcommand{\epf}{$\quad$\hfill
\raisebox{0.11truecm}{\fbox{}}\par\vskip0.4truecm}

\usepackage{array}
\newcolumntype{M}[1]{>{\centering\arraybackslash}m{#1}}
\newcolumntype{R}[1]{>{\raggedleft\arraybackslash}m{#1}}
\newcolumntype{N}{@{}m{0pt}@{}}

\begin{document}

\title {Generating series for GUE correlators}
\author
{
{Boris Dubrovin, Di Yang}\\
{\small SISSA, via Bonomea 265, Trieste 34136, Italy}\\
}
\date{}
\maketitle
\begin{abstract}
We extend to the Toda lattice hierarchy the  approach of \cite{BDY1, BDY2} to computation of logarithmic derivatives of tau-functions in terms of the so-called \textit{matrix resolvents} of the corresponding difference Lax operator.
As a particular application we 
obtain explicit generating series for connected GUE correlators. On this basis an efficient recursive procedure for computing the correlators in full genera is developed.
\end{abstract}

\section{Introduction}

\setcounter{equation}{0}
\setcounter{theorem}{0}
\subsection{Formulation of the main result}\par

Denote ${\mathcal H}(N)$ the space of $N\times N$ Hermitean matrices.  
The Gaussian Unitary Ensemble (GUE) correlators of observables $\tr \, M^i$,  $i=1, \, 2, \dots$ with respect to the Gaussian probability measure on ${\mathcal H}(N)$ are defined by
\beq\label{glav1}
\left\langle  \tr\, M^{i_1}\cdots \tr\, M^{i_k} \right\rangle:=\frac{\int\limits_{{\mathcal H}(N)}  \tr\, M^{i_1}\dots \tr\, M^{i_k}\,e^{-\frac12 \tr\, M^2} \, dM}{\int\limits_{{\mathcal H}(N)}  e^{-\frac12 \tr\, M^2}dM}.
\eeq
They are certain polynomials in $N$ that can be computed by the Wick rule. By $\left\langle  \tr\, M^{i_1}\cdots \tr\, M^{i_k} \right\rangle_c$
we denote the \emph{connected} correlators. That means that, applying the Wick rule to the computation of Gaussian integrals \eqref{glav1} we keep summation over connected Feynman diagrams only. For example,
$\langle \, \left(\tr\,M^2\right)^2 \, \rangle= 2N^2 + N^4$ but $\langle \, \left(\tr\,M^2\right)^2 \, \rangle_c= 2N^2$;  see more details in Appendix \ref{appa}. According to \cite{thooft1, thooft2, BIZ} the connected GUE correlators have an important application to the problem of enumeration of ribbon graphs on two-dimensional oriented surfaces, see details in Appendix A below. This was one of the motivations for a significant interest to the problem of computation of the GUE correlators, see e.g. \cite{BI,  fle, KKN, Pierce}.

For every $k\geq 1$ we will consider generating series of the $k$-point correlators of the form
\beq\label{glav2}
C_k(N;\lambda_1, \dots, \lambda_k):=\sum_{i_1, \dots, i_k=1}^\infty 
\frac{\left\langle  \tr\, M^{i_1}\cdots \tr\, M^{i_k} \right\rangle_c}{\lambda_1^{i_1+1}\dots \lambda_k^{i_k+1}}
\eeq
where $\lambda_1$, \dots, $\lambda_k$ are independent variables, $N$ refers to the size of the Hermitean matrices. Our main result is the following explicit expressions for the generating series \eqref{glav2}.

\begin{theorem} \label{thm1} 1) {The generating series for $1$-point} correlators has the form
\beq\label{onepoint}
C_1(N;\lambda)= {N} \, \sum_{j\geq 0} \frac{(2j-1)!!}{\lambda^{2j+1}} \left[ {}_2F_1(-j,-N; 2; 2) -j \cdot {}_2F_1(1-j, 1-N; 3; 2)\right].
\eeq
2) Introduce a $2\times 2$ matrix-valued series
\beq\label{ram}
{\mathcal R}_n(\lambda):=\left(\begin{array}{cc} 1 & 0\\ 0 & 0\end{array}\right)+n\sum_{j=0}^\infty \frac{(2j-1)!!}{\lambda^{2j+2}} \left(\begin{array}{cc} (2j+1)A_{n,j} & -\lambda\,  B_{n+1,j}\\
\\
\frac{\lambda}{n}\, B_{n,j} & -(2j+1) A_{n,j}\end{array}\right)\in {\rm Mat}\left(2, \mathbb Z[n] \left[[\lambda^{-1}]\right] \right)
\eeq
where
\eqa\label{ank}
&&
A_{n,j}=\frac1{n}\sum_{i=0}^j 2^i \left( \begin{array}{c} j \\ i \end{array}\right)\left(\begin{array}{c}n\\i+1\end{array}\right) ={}_2F_1(-j,1-n; 2; 2)
\nn\\
&&
\\
&&
B_{n,j}=\sum_{i=0}^j 2^i \left( \begin{array}{c} j \\ i \end{array}\right)\left(\begin{array}{c}n-1\\i\end{array}\right)={}_2F_1(-j,1-n; 1; 2).
\nn
\eeqa
Then
\beq\label{twopoint}
C_2(N; \lambda_1,\lambda_2)=\frac{\tr\, {\mathcal R}_N(\lambda_1) {\mathcal R}_N(\lambda_2)-1}{(\lambda_1-\lambda_2)^2}
\eeq

\beq\label{kpoint}
C_k(N; \lambda_1, \dots, \lambda_k)=-\frac1{k}\sum_{\sigma\in S_k} \frac{\tr \,\left[{\mathcal R}_N(\lambda_{\sigma_1})\dots {\mathcal R}_N(\lambda_{\sigma_k})\right]}{(\lambda_{\sigma_1}-\lambda_{\sigma_2})\dots (\lambda_{\sigma_{k-1}}-\lambda_{\sigma_k}) (\lambda_{\sigma_k}-\lambda_{\sigma_1})}, \quad k\geq 3.
\eeq 
\end{theorem}

In the above formulae
$$
{}_2F_1(a,b; c; z)=\sum_{j=0}^\infty \frac{(a)_j (b)_j}{(c)_j} \frac{z^j}{j!}=1+\frac{a\, b}{c} \frac{z}{1!} + \frac{a(a+1)\, b(b+1)}{c(c+1)}\frac{z^2}{2!}+\dots
$$
is the Gauss hypergeometric function. Recall that it truncates to a polynomial if $a$ or $b$ are non-positive integers.

\begin{remark} To the best of our knowledge the first generating series for the one-point connected correlators was obtained by J.~Harer and D.~Zagier in \cite{HZ}. In \cite{MS} A.~Morozov and Sh.~Shakirov constructed a generating function of the Harer--Zagier type for the two-point correlators.  These generating functions are different from ours but produce identical results for the correlators.
\end{remark}

\setcounter{equation}{0}
\setcounter{theorem}{0}
\subsection{Matrix resolvent of a second order difference operator \\ and tau-function of the Toda lattice}\par 

Consider a second order difference operator $L$ acting on functions $\psi_n$, $n\in\mathbb Z$ by
\begin{equation}\label{lax0}
\left( L\, \psi\right)_n =\psi_{n+1}+v_n \, \psi_n +w_n  \,  \psi_{n-1}.
\end{equation}
The standard realization of the Toda lattice hierarchy is given by the Lax representation
\begin{eqnarray}\label{toda1}
&&
\frac{\partial L}{\partial t_j} =\left[ A_j,L\right], \quad j\geq 0
\\
&&
A_j =\left(L^{j+1}\right)_+.
\label{toda2}
\end{eqnarray}
Introduce the matrix
\begin{equation}\label{mlax}
U_n(\lambda)= \left(\begin{array}{cc} v_n-\lambda & w_n\\ -1 & 0\end{array}\right).
\end{equation}

Observe that the second order difference equation for the eigenfunctions of the Lax operator
$$
L\, \psi =\lambda\, \psi
$$
can be written in the matrix form
\begin{equation}\label{mlax1}
\Delta\, \Psi_n +U_n(\lambda) \, \Psi_n=0, \quad \Psi_n =\left(\begin{array}{c} \psi_n\\ \psi_{n-1}\end{array}\right)
\end{equation}
where $\Delta$ is the shift operator
$$
\Delta\, \Psi_n = \Psi_{n+1}.
$$

Introduce the ring $\mathbb Z[{\bf v}, {\bf w}]$ of polynomials with integer coefficients in the infinite set of variables ${\bf v}=(v_n)$, ${\bf w}=(w_n)$, $n\in\mathbb Z$.

\begin{lemma} \label{lemmaone} There exists a unique $2\times 2$ matrix series 
$$
R_n(\lambda)=\left(\begin{array}{cc}1 & 0\\0 & 0\end{array}\right)+{\mathcal O}\left(\lambda^{-1}\right)\in {\rm Mat} \left(2, \mathbb Z[{\bf v}, {\bf w}] \left[[\lambda^{-1}]\right]\right)
$$ 
satisfying equation
\beq\label{eqres}
R_{n+1}(\lambda) U_n(\lambda)-U_n(\lambda) R_n(\lambda)=0
\eeq
along with the normalization conditions
\beq\label{normres}
\tr \,R_n(\lambda)=1, \quad \det R_n(\lambda)=0.
\eeq
\end{lemma}

\begin{defi} The series $R_n(\lambda)$ is called the \emph{matrix resolvent} of the difference operator $L$.
\end{defi}

Let the difference operator $L$ depend on the times of the Toda lattice hierarchy \eqref{toda1}, \eqref{toda2}. Then so does the matrix resolvent, $R_n=R_n({\bf t},\lambda)$. Here and below ${\bf t}:=(t_0,t_1,\dots)$. We will now present an algorithm for computing the so-called tau-function of an arbitrary solution $v_n({\bf t})$, $w_n({\bf t})$. 
The tau-function of a solution will be determined uniquely, up to a simple factor, by an explicitly written collection of its second order logarithmic derivatives in the continuous variables $t_k$ and the discrete variable $n$. 

\begin{lemma} \label{lemmatwo} For any solution $v_n=v_n({\bf t})$, $w_n=w_n({\bf t})$ to the Toda lattice hierarchy there exists a function $\tau_n({\bf t})$ such that
\eqa\label{taun1}
&&
\sum_{i,\, j\geq 0} \frac{1}{\lambda^{i+2} \mu^{j+2}} \frac{\partial^2\log\tau_n({\bf t})}{\partial t_i \,\partial t_j}=\frac{{\rm tr}\, R_n({\bf t},\lambda)R_n({\bf t},\mu)-1}{(\lambda-\mu)^2}
\\
&&\label{taun2}
\sum\limits_{i\geq 0} \frac1{\lambda^{i+2}}\frac{\pal}{\pal t_i} \log \frac{\tau_{n+1}({\bf t})}{\tau_n({\bf t})}  =  \left[ R_{n+1}({\bf t},\lambda)\right]_{21}
\\
&&\label{taun3}
\frac{\tau_{n+1}({\bf t}) \tau_{n-1}({\bf t})}{\tau_n^2({\bf t})}=w_n.
\eeqa
The function $\tau_n({\bf t})$ is determined uniquely by the solution $v_n({\bf t})$, $w_n({\bf t})$ up to 
$$
\tau_n({\bf t}) \mapsto e^{a_0 + a_1 n + \sum_{j\geq 0} b_j t_j} \tau_n({\bf t})
$$
for some constants $a_0$, $a_1$, $b_0$, $b_1$, $\dots$ independent of $n$.
\end{lemma}


\begin{defi} \label{defitau} 
The function $\tau_n({\bf t})$ defined by eqs. \eqref{taun1}--\eqref{taun3} is called the \emph{tau-function} of the solution $v_n({\bf t})$, $w_n({\bf t})$.
\end{defi}

Remarkably, the higher logarithmic derivatives of the tau-function can also be expressed in terms of the matrix resolvent.

\begin{theorem} \label{thm2} The order $k\geq 3$ logarithmic derivatives of the tau-function of a solution to the Toda lattice hierarchy can be computed from the following generating series
\beq\label{mult-pt}
\sum_{i_1, \dots, i_k=0}^\infty \frac1{\lambda_1^{i_1+2}\dots \lambda_k^{i_k+2}} \frac{\pal^k\log\tau_n({\bf t})}{\pal t_{i_1}\dots \pal t_{i_k}}=-\frac1{k}\sum_{\sigma\in S_k} \frac{\tr \,\left[{ R}_n({\bf t},\lambda_{\sigma_1})\dots { R}_n({\bf t},\lambda_{\sigma_k})\right]}{(\lambda_{\sigma_1}-\lambda_{\sigma_2})\dots (\lambda_{\sigma_{k-1}}-\lambda_{\sigma_k}) (\lambda_{\sigma_k}-\lambda_{\sigma_1})}.
\eeq 
\end{theorem}

It is well-known \cite{Witten}, \cite{gmmmo} that the GUE partition function (see eq. \eqref{part} below) is the tau-function of a particular solution to the Toda lattice, identifying $s_k=t_{k-1}$. Logarithmic derivatives of the tau-function evaluated at ${\bf t}=0$ coincide with the connected correlators \eqref{glav1}. This solution is specified by the initial data
\beq\label{init}
v_n({\bf t}=0)=0, \quad w_n({\bf t}=0)=n.
\eeq
Hence it remains to compute the matrix resolvent of the operator
\beq\label{init1}
\Delta-\left( \begin{array}{cr}\lambda & -n\\ 1 & 0\end{array}\right).
\eeq

\begin{theorem} \label{thm3} The matrix resolvent of the operator \eqref{init1} coincides with series ${\mathcal R}_n(\lambda)$ given by eq. \eqref{ram}.
\end{theorem}

Theorem \ref{thm1} readily follows from Theorems \ref{thm1} and \ref{thm2}.

\medskip

The approach of the present paper can be generalized to the integrable systems associated with higher order difference Lax operators. Such a generalization will be developed in a subsequent publication.

\paragraph{Organization of the paper.} In Sect.\,\ref{s2} we prove  Lem.\,\ref{lemmaone}, Lem.\,\ref{lemmatwo} and Thm.\,\ref{thm2}.
In Sect.\,\ref{s3} we prove Thm.\,\ref{thm3} and give an algorithm of computing connected GUE correlators in a recursive way. 
In Sect.\,\ref{s4} we outline an algorithm for computing the genus expansion of the GUE free energy based on \cite{DZ,Du1,icmp}.
A short review on the Hermitean matrix model, mainly following \cite{BIZ, mehta, gmmmo}, is given in Appendix \ref{appa}.

\paragraph{Acknowledgements}
The work is supported by PRIN 2010-11 Grant ``Geometric and analytic theory of 
Hamiltonian systems in finite and infinite dimensions" of Italian Ministry of Universities and Researches. 

\section{Computing tau-functions of Toda lattice hierarchy} \label{s2}  \par

\setcounter{equation}{0}
\setcounter{theorem}{0}
\subsection{Matrix resolvents. Proof of Lemma \ref{lemmaone}}\par

In this section we will remind basic constructions of the Toda lattice hierarchy. We will also prove the Lemma \ref{lemmaone}.

The Toda lattice is a system of particles on the line with exponential interaction of neighbors. The Hamiltonian is written as a formal infinite sum
$$
H(q,p) =\sum_{n\in\mathbb Z} \frac{p_n^2}2 + e^{q_n-q_{n+1}}.
$$
After the substitution
\beq\label{substi}
v_n=-\dot q_n, \quad w_n =e^{q_{n-1}-q_n}
\eeq
the equations of motion
$$
\ddot q_n =e^{q_{n-1}-q_n}-e^{q_n-q_{n+1}}, \quad n\in\mathbb Z
$$
take the form
\beq\label{toda0}
\begin{array}{rcl} \dot v_n & = & w_{n+1} - w_n\\
\dot w_n & = & w_n (v_n-v_{n-1}).\end{array}
\eeq
Eqs. \eqref{toda0} are considered as a differential-difference evolution system with the time variable $t$ and discrete spatial variable $n\in\mathbb Z$.
Integrability of the Toda equations was discovered by H. Flaschka \cite{fla} and S. Manakov \cite{mana}. The corresponding Lax operator is a second order difference operator \eqref{lax0}. The standard realization of the commuting flows of the Toda lattice hierarchy is given by the Lax representation \eqref{toda1}.
The first flow of the hierarchy coincides with \eqref{toda0}, $t=t_0$, then
$$
\frac{\partial v_n}{\partial t_1}=w_{n+1}(v_{n+1}+v_n) -w_n(v_n+v_{n-1}), \quad \frac{\partial w_n}{\partial t_1} =w_n (w_{n+1}-w_{n-1} +v_n^2 - v_{n-1}^2)
$$
etc. They are Hamiltonian equations of the form
$$
\frac{\pal v_n}{\pal t_j} = \{ v_n, H_j\},\qquad \frac{\pal w_n}{\pal t_j} = \{w_n, H_j\},\qquad \quad j\geq 0
$$
 with respect to the Poisson brackets
\begin{equation}\label{bra}
\{ q_n, p_m\} =\delta_{mn} \quad \Rightarrow\quad \{ v_n, w_n\} =-w_n, \quad \{ v_n, w_{n+1}\}=w_{n+1}
\end{equation}
(use the substitution \eqref{substi}) with the Hamiltonians
\begin{equation}\label{ham}
H_j =\sum_n h_j(n), \quad h_j(n)=\frac1{j+2}\,\left( L^{j+2} \right)_{nn}, \qquad j\geq -1.
\end{equation}
Here $\left( L^{j+2} \right)_{nn}$ means taking the $n$-th diagonal entry of the infinite matrix $L^{j+2}$. The first several Hamiltonian ``densities" read
$$
h_{-1}(n)=v_n,\qquad h_0(n)=\frac12(v_n^2 + w_n + w_{n+1})
$$
$$
h_1(n)=\frac13[v_n^3 + 2 v_n (w_n+w_{n+1}) + w_n v_{n-1} +w_{n+1} v_{n+1}].
$$
Note that $h_{-1}(n)$ is the density of one of the Casimirs of the Poisson bracket. 
The Hamiltonian densities $h_j(n),\, j\geq -1$  satisfy
$$
(i+1) \frac{\pal h_{j-1} }{ \pal t_i}  = (j+1) \frac{\pal h_{i-1}}{\pal t_j},\qquad \forall\,i,j\geq 0.
$$
Observe that, changing the normalization 
$$
\tilde h_j =\frac1{(j+1)!} h_j\quad\Rightarrow\quad \tilde t_j=(j+1)! \, t_j
$$
one arrives at the \textit{tau-symmetry} property
\beq\label{tausym}
\frac{\pal \tilde h_{j-1}}{\pal \tilde t_i}=\frac{\pal \tilde h_{i-1}}{\pal \tilde t_j}.
\eeq
Such a normalization of the Hamiltonians/time variables of the hierarchy is used when working with the Gromov--Witten invariants of ${\bf P}^1$ \cite{DZ-toda,CDZ}.

It will be more convenient to work with the $\lambda$-dependent first order matrix version \eqref{mlax1}, \eqref{mlax} of the Lax operator acting on two-component vector-valued functions on the lattice
$$
\Psi_n=\left(\begin{array}{c}\psi_n\\\psi_{n-1}\end{array}\right).
$$
Notice that the equation \eqref{eqres} for what we call matrix resolvent $R_n(\lambda)$ can be written in the form
\beq\label{rescom}
\left[\Delta+U_n(\lambda), R_n(\lambda)\right]=0.
\eeq
Clearly the normalization conditions \eqref{normres} are compatible with eq. \eqref{eqres}.

Let us proceed with the proof of Lemma \ref{lemmaone}. Write
$$
R_n(\lambda)=\left(   \begin{array}{cc} 1+ \alpha_n(\lambda) & \beta_n(\lambda) \\ \gamma_n(\lambda) & -\alpha_n(\lambda) \end{array} \right).
$$
Substituting this expression in \eqref{eqres} we obtain
\eqa
&& \beta_n = -w_n \,\gamma_{n+1} \label{rr1} \\
&& \alpha_{n+1} +\alpha_n+1= \gamma_{n+1}  \, (\lambda-v_n) \label{rr2} \\
&& (\lambda-v_n) (\alpha_n-\alpha_{n+1}) = w_n\, \gamma_n - w_{n+1} \, \gamma_{n+2}. \label{rr3} 
\eeqa
Expand  
$$
\gamma_n= \sum_{j\geq 0} \frac{c_{n,j}}{\lambda^{j+1}},\qquad \alpha_n=   \sum_{j\geq 0} \frac{a_{n,j}}{\lambda^{j+1}}.
$$
It follows immediately from \eqref{rr2}--\eqref{rr3} the recursion relations:
\eqa
&& c_{n,j+1} = v_{n-1} \, c_{n,j} + a_{n,j} +a_{n-1,j}.\label{rec1}\\
&& a_{n,j+1}-a_{n+1,j+1} + v_n \, [ a_{n+1,j}-a_{n,j}]  + w_{n+1} \,c_{n+2,j}- w_n \, c_{n,j} =0. \label{rec2} 
\eeqa
The normalization conditions \eqref{normres} imply  
$$
a_{n,0}=0, \qquad c_{n,0}=1
$$
along with another recursion relation
\beq\label{rec3}
a_{n,\ell}=w_n\left( c_{n+1,\ell-1}+c_{n,\ell-1}\right)+\sum_{i+j=\ell-1} \left[ a_{n,i} a_{n,j} + c_{n,i} c_{n+1,j}\right] 
\eeq
The Lemma \ref{lemmaone} readily follows from the recursion relations \eqref{rec1} and \eqref{rec3}. 

\setcounter{equation}{0}
\setcounter{theorem}{0}
\subsection{Matrix resolvents and Toda flows. Proof of Lemma \ref{lemmatwo} and Thm.\,\ref{thm3}} \par

We will now represent equations of the Toda flows in terms of the matrix resolvent $R_n$.  

\begin{lemma} \label{flow-ac}
The Toda flows \eqref{toda1} can be written in terms of $c_{n,j},\,a_{n,j}$ as follows:
\eqa
&& \frac{\pal v_n}{\pal t_j} = a_{n+1,j+1} - a_{n,j+1},\qquad j\geq 0 \label{v-flow}\\
&& \frac{\pal w_n} {\pal t_j} = w_n \, \left(c_{n+1,j+1}-c_{n,j+1}\right),\qquad j\geq 0. \label{w-flow}
\eeqa
\end{lemma}
\noindent \textit{Proof}. By using the recursion relations \eqref{rec1},\,\eqref{rec2} and by comparing them with \eqref{toda1}.\epf

Lemma \ref{flow-ac} implies the following
\begin{cor}
\label{tau-h1} 
For any $j\geq -1$, the following formula holds true
$$
h_j(n)= \frac{1}{j+2} c_{n+1,j+2}.
$$
\end{cor}
\begin{lemma} The functions $\alpha_n=\alpha_n(\lambda),\gamma_n=\gamma_n(\lambda)$ satisfy
\eqa \label{ham}
&& \!\!\!\!\!\!\!\! \gamma_{n+3}=
 \frac{ \gamma_{n+2}  } {w_{n+2}} (\lambda -v_{n+1})^2+ \frac{ (\lambda -v_{n+1})(\gamma_{n} w_n-\gamma_{n+2} w_{n+1})}{w_{n+2} \, (\lambda -v_n)}+\frac{\gamma_{n+1} \,  \left[w_{n+1}-(\lambda -v_n) (\lambda -v_{n+1})\right]}{w_{n+2} } \nn\\
&& \\
&&\!\!\!\!\!\!\!\! (\lambda-v_{n-1})  \left[w_{n+1} (\alpha_{n+2}+\alpha_{n+1}+1)-(\lambda-v_n)(\lambda-v_{n+1}) \alpha_{n+1}\right]  \nn\\
&&\!\!\!\!\!\!\!\! \qquad\qquad =(\lambda-v_{n+1})  \left[w_n (\alpha_{n}+\alpha_{n-1}+1)-(\lambda-v_n)(\lambda-v_{n-1}) \alpha_{n}\right].
\eeqa
\end{lemma}
\noindent \textit{Proof}. \quad By using \eqref{rr2}--\eqref{rr3} and by eliminating one of the series $\alpha_n(\lambda)$, $\gamma_n(\lambda)$.  \epf

For any $j\geq 0$, define the matrix-valued function
$$
V_{n,j}(\lambda)= \left[\lambda^{j+1} R_n(\lambda)\right]_+ +\left( \begin{array}{cc} 0 & 0\\ 0 & c_{n,j+1}\\ \end{array}\right)
$$
where $[~~]_+$ means taking the polynomial part in $\lambda$.
The flows of the Toda lattice hierarchy can be represented as the following Lax equation
$$
\frac{\partial U_n(\lambda)} { \partial t_j} = V_{n+1,j}(\lambda) \, U_n(\lambda) - U_n(\lambda) \, V_{n,j}(\lambda), \qquad j\geq 0
$$
which are the compatibility conditions between eq.\,\eqref{mlax1} and 
\beq\label{psiT}
\frac{\partial \Psi_n} {\partial t_j} = V_{n,j}(\lambda) \, \Psi_n,\qquad j=0,1,2,\dots.
\eeq

Introduce an operator  $\nabla(\lambda)$ depending on a parameter $\lambda$ by
\beq\label{nabladef}
\nabla(\lambda):= \sum_{j\geq 0} \frac1{\lambda^{j+2}}\frac{\pal}{\pal t_j}.
\eeq
From eq.\,\eqref{psiT} it readily follows that 
$$
\nabla(\mu) \, \Psi_n(\lambda) = \left[\frac{R_n(\mu)}{\mu-\lambda} + Q_n(\mu)\right] \, \Psi_n(\lambda)
$$
where $$Q_n(\mu):= -\frac{\rm id}{\mu} + \left( \begin{array}{cc} 0 & 0 \\ 0 & \gamma_n(\mu) \\ \end{array} \right).$$
We arrive at
\begin{lemma} 
The following equation holds true
\beq\label{naR}
\nabla(\mu) \, R_n(\lambda) = \frac1{\mu-\lambda} \left[ R_n(\mu), R_n(\lambda) \right] + [Q_n(\mu),R_n(\lambda)].
\eeq
\end{lemma}

We are now in a position to prove Lemma \ref{lemmatwo}. 

\noindent \textit{Proof} of Lemma \ref{lemmatwo}. \quad 
First, let us check that
\beq\label{numres}
\tr \, R_n(\lambda) R_n(\mu)-1
\eeq
is divisible by $(\lambda-\mu)^2$. Indeed, from the normalization 
conditions \eqref{normres} it readily follows that $\tr\, R_n^2(\lambda)=1$. So \eqref{numres} is 
divisible by $(\lambda-\mu)$. Due to symmetry in $\lambda$, $\mu$ this implies divisibility by $(\lambda-\mu)^2$. 
Thus the  r.h.s. of the first eq. in \eqref{taun1} is a formal series in negative powers of $\lambda,\,\mu$,  
$$
\frac{\tr \, R_n(\lambda)R_n(\mu) -1 }{(\lambda-\mu)^2} = \sum_{i,j\geq 0} \frac{ \Omega_{i;j}(n) }{\lambda^{i+2} \, 
\mu^{j+2}}\quad \mbox{for some coefficients}\quad\Omega_{i;j} (n)\in \mathbb Z[{\bf v}, {\bf w}].
$$
The first few of them are
$$
\Omega_{0; 0}(n)=w_n, \quad \Omega_{0; 1}(n)=w_n(v_n+v_{n-1}), \quad \Omega_{1;1}(n)=w_n\left[ w_{n+1}+w_{n-1}+(v_n+v_{n-1})^2\right].
$$
Clearly
$$
\Omega_{i;j}=\Omega_{j;i},\qquad \forall\, i,j\geq 0.
$$ 

Let us compute the time-derivatives of these coefficients. We have
\eqa
\sum_{k,\ell,m\geq 0}  \frac{\pal_{t_m} \, \Omega_{k;\ell} }{ \nu^{m+2} \lambda^{k+2} \mu^{\ell+2}} 
&=& \nabla(\nu)\,  \sum_{k,\ell\geq 0}  
\frac{\Omega_{k;\ell}}{\lambda^{k+2} \mu^{\ell+2}} 
\nn\\
&= & \frac{  \tr \left ( R(\mu) \, \nabla(\nu) R (\lambda) \right)}{(\lambda-\mu)^2}  
+ \frac{  \tr \left(R (\lambda) \, \nabla(\nu)R(\mu)\right)}{(\lambda-\mu)^2} \nn\\
&= & \frac{ \tr \left([R(\nu),R(\lambda)] \, R(\mu)\right)}{(\lambda-\mu)^2(\nu-\lambda)}  - 
 \frac{  \tr \left([Q(\nu) ,R(\lambda)] \, R(\mu)\right)}{(\lambda-\mu)^2} \nn\\
&& + \frac{  \tr \left(R (\lambda) \, [R(\nu),R(\mu)]\right)}{(\lambda-\mu)^2(\nu-\mu)}
-  \frac{  \tr \left(R (\lambda) \, [Q(\nu),P(\mu)]\right)}{(\lambda-\mu)^2}\nn \\
&=&  - \frac{\tr \left([R(\nu),R(\lambda)] \, R(\mu)\right)} {(\lambda-\mu)(\mu-\nu)(\nu-\lambda)}. \nn
\eeqa
It is easy to see that the last expression is symmetric in $\lambda$, $\mu$, $\nu$.
Hence  
$$
\frac{ \pal \Omega_{k;\ell}}{\pal {t_m}} = \frac{\pal \Omega_{m;\ell}}{\pal {t_k} }\quad \forall\, k,l,m\geq0.
$$

We are now to prove compatibility between eq.\,\eqref{taun2} and eq.\,\eqref{taun1}.  On one hand,
we have
\eqa 
&& \sum_{i,\, j\geq 0} \frac{1}{\lambda^{i+2} \mu^{j+2}} \left[ \Omega_{i;j}(n+1)-\Omega_{i;j}(n) \right] \nn\\
& =& \frac{\tr \, R_{n+1}(\lambda) R_{n+1}(\mu) - \tr \, R_n(\lambda) R_n(\mu)}{(\lambda-\mu)^2}\nn\\
 &=& \frac{(1+ 2 \alpha_{n} (\lambda)) \, \gamma_{n}(\lambda) - (1+2\alpha_n(\mu)) \, \gamma_{n+1}(\lambda)}{\lambda-\mu} - \gamma_{n+1}(\lambda)\, \gamma_{n+1}(\mu)\nn\\
 &=& \frac{(1+ 2 \alpha_{n+1} (\mu)) \, \gamma_{n+1}(\lambda) - (1+2\alpha_{n+1}(\lambda)) \, \gamma_{n+1}(\mu)}{\lambda-\mu} + \gamma_{n+1}(\lambda)\, \gamma_{n+1}(\mu) \nn
\eeqa
where the last equality uses the relation \eqref{rr2}. 
On another hand, it follows from eq.\,\eqref{naR} that 
\beq\label{e2}
\nabla(\mu) \gamma_{n+1}(\lambda)= \frac{\gamma_{n+1}(\lambda)(1+2\alpha_{n+1}(\mu))-\gamma_{n+1}(\mu)(1+2\alpha_{n+1}(\lambda))}{\lambda-\mu} +\gamma_{n+1}(\lambda)\gamma_{n+1}(\mu).
\eeq
Hence 
$$
\sum_{i,\, j\geq 0} \frac{1}{\lambda^{i+2} \mu^{j+2}} \left[ \Omega_{i;j}(n+1)-\Omega_{i;j}(n) \right] =  \nabla(\mu) \gamma_{n+1}(\lambda).
$$

Finally we show the compatibility between eq.\,\eqref{taun3} and eqs.\,\eqref{taun1},\,\eqref{taun2}.
Indeed,
\eqa
&& \sum_{i,\, j\geq 0} \frac{1}{\lambda^{i+2} \mu^{j+2}} \left[ \Omega_{i;j}(n+1) + \Omega_{i;j}(n-1)-2 \, \Omega_{i;j}(n) \right] \nn\\
&=& \frac{\tr \, R_{n+1}(\lambda) R_{n+1}(\mu) +\tr \, R_{n-1}(\lambda) R_{n-1}(\mu) - 2 \, \tr \, R_n(\lambda) R_n(\mu)}{(\lambda-\mu)^2}\nn\\
&=& \frac{(1+ 2 \alpha_{n+1} (\mu)) \, \gamma_{n+1}(\lambda) - (1+2\alpha_{n+1}(\lambda)) \, \gamma_{n+1}(\mu)}{\lambda-\mu} + \gamma_{n+1}(\lambda)\, \gamma_{n+1}(\mu)\nn\\
&& -\frac{(1+ 2 \alpha_{n} (\mu)) \, \gamma_{n}(\lambda) - (1+2\alpha_{n}(\lambda)) \, \gamma_{n}(\mu)}{\lambda-\mu} - \gamma_{n}(\lambda)\, \gamma_{n}(\mu). \nn
\eeqa
Also,
\eqa
\nabla(\mu)\nabla(\lambda) \, \log w_n
&=& \nabla(\mu) \left[ \gamma_{n+1}(\lambda)- \gamma_n(\lambda)-\lambda^{-1}\right]  \nn\\
&=&  \left[\frac{\gamma_{n+1}(\lambda)(1+2\alpha_{n+1}(\mu))-\gamma_{n+1}(\mu)(1+2\alpha_{n+1}(\lambda))}{\lambda-\mu} +\gamma_{n+1}(\lambda)\gamma_{n+1}(\mu)\right] \nn\\
&&- \left[\frac{\gamma_{n}(\lambda)(1+2\alpha_{n}(\mu))-\gamma_{n}(\mu)(1+2\alpha_{n}(\lambda))}{\lambda-\mu} +\gamma_{n}(\lambda)\gamma_{n}(\mu)\right].\nn
\eeqa
Hence 
$$
\sum_{i,\, j\geq 0} \frac{1}{\lambda^{i+2} \mu^{j+2}} \left[ \Omega_{i;j}(n+1) + \Omega_{i;j}(n-1)-2\Omega_{i;j}(n) \right] = \nabla(\mu)\nabla(\lambda) \, \log w_n.
$$
This proves compatibility between \eqref{taun3} and \eqref{taun1}. 
The compatibility between \eqref{taun3} and \eqref{taun2} is equivalent to eq.\,\eqref{w-flow}.

As a result, for an arbitrary solution $v_n({\bf t}),w_n({\bf t})$ to the Toda lattice hierarchy, 
there exists a function $\tau_n({\bf t})$ of this solution satisfying \eqref{taun1}--\eqref{taun3}. 
It is easy to see that the freedom of 
$\tau_n$ satisfying \eqref{taun1}--\eqref{taun3} is only an arbitrary factor of
the form
$$
e^{a_0 + a_1 n + \sum_{j\geq 0} b_j t_j}
$$
where $a_0,a_1,b_0,b_1,b_2,\dots$ are constants independent of $n$.
The lemma is proved.
\epf

\begin{defi} For a given $k\geq 2$ and a given set of integers $i_1,\dots,i_k\geq 0$, we call 
$$ \frac{\pal^k\log\tau_n({\bf t})}{\pal t_{i_1}\dots \pal t_{i_k}}$$ the $k$-point correlation functions of the given solution $v_n({\bf t})$, $w_n({\bf t})$ of the Toda lattice hierarchy. 
\end{defi}

Also the first logarithmic derivatives (``one-point correlation functions") of the tau-function will be under consideration. They are determined by a solution $v_n({\bf t})$, $w_n({\bf t})$ up to additive constants.

Clearly the above proof of Lemma \ref{lemmatwo} already shows that
\begin{cor} \label{cor225} The generating series of three-point correlation functions of an arbitrary solution to the Toda lattice hierarchy has the following expression
$$
\sum_{i,j,l\geq 0} \frac{\pal_{t_i}\pal_{t_j}\pal_{t_l} \log \tau_n}{\lambda^{i+2} \mu^{j+2}\nu^{l+2}} =  - \frac{\tr \, R_n(\lambda) R_n(\mu) R_n(\nu) - \tr \, R_n(\lambda) R_n(\nu) R_n(\mu)  } {(\lambda-\mu)(\mu-\nu)(\nu-\lambda)}.
$$
\end{cor}

\noindent \textit{Proof} of Thm.\,\ref{thm2}   \quad 
For any permutation $\sigma=[\sigma_1,\dots,\sigma_q]\in S_q,\,q\geq 2,$ define
$$
P(\sigma):= -\prod_{j=1}^q \frac{1}{\lambda_{\sigma_j}-\lambda_{\sigma_{j+1}}}, \qquad \sigma_{q+1}:=\sigma_1.
$$
We use mathematical induction for the proof. For $k=3$, the formula \eqref{kpoint} is already obtained in Corollary \ref{cor225}.
Suppose \eqref{kpoint} is true for $k=p,\,p\geq 3$. Then for $k=p+1$, we have\footnote{In this calculation we omit the index $n$ of $\tau_n$ and $R_n$.}
\eqa
&&  \sum_{i_1, \dots, i_{p+1}=0}^\infty \frac1{\lambda_1^{i_1+2}\dots \lambda_{p+1}^{i_{p+1}+2}} \frac{\pal^k\log\tau({\bf t})}{\pal t_{i_1}\dots \pal t_{i_{p+1}}}  =
 - \frac {1} {p } \, \nabla(\lambda_{p+1})\,  \sum_{\sigma \in S_p} 
\frac { \tr\, \left(R(\lambda_{\sigma_1})\dots R(\lambda_{\sigma_p})\right)}{\prod_{j=1}^{p} (\lambda_{\sigma_j} - \lambda_{\sigma_{j+1}})}\nn\\
&=&- \frac {1} {p} \,  \sum_{\sigma \in S_p} \sum_{q=1}^p
\frac { \tr\,
\left(R(\lambda_{\sigma_1}) \dots  
\left[\frac{R(\lambda_{p+1})}{\lambda_{p+1} - \lambda_{\sigma_q}}  + Q(\lambda_{p+1}) ,R(\lambda_{\sigma_q})\right] \dots R(\lambda_{\sigma_p})\right)}{\prod_{j=1}^{p} (\lambda_{\sigma_j} - \lambda_{\sigma_{j+1}})}\nn\\
&=& \frac {1} {p} \,  \sum_{\sigma \in S_p} P(\sigma) \sum_{q=1}^p (\lambda_{\sigma_q}-\lambda_{\sigma_{q-1}})\frac { 
\tr\, R(\lambda_{p+1}) R(\lambda_{\sigma_q}) \dots R(\lambda_{\sigma_p}) R(\lambda_{\sigma_1}) \dots R(\lambda_{\sigma_{q-1}})}{(\lambda_{p+1}-\lambda_{\sigma_q})(\lambda_{p+1}-\lambda_{\sigma_{q-1}})}\nn\\
&=&\frac1 {p} \sum_{q=1}^p \,  \sum_{\sigma \in S_p} P([p+1,s_q,\dots,s_p,s_1,\dots,s_{q-1}])\,\tr\,R(\lambda_{p+1}) R(\lambda_{\sigma_q}) \dots R(\lambda_{\sigma_p}) R(\lambda_{\sigma_1}) \dots R(\lambda_{\sigma_{q-1}})\nn\\
&=& \sum_{\sigma \in S_p} P([p+1,\sigma])
\, \tr\, R(\lambda_{p+1}) R(\lambda_{\sigma_1}) \dots R(\lambda_{\sigma_p}).\nn
\eeqa
The theorem is proved. \epf

The resulting expressions for the generating series can be used for developing efficient algorithms for computing the GUE correlators. To this end it is convenient to represent the multipoint formula \eqref{kpoint} in a slightly modified way.

\begin{cor} \label{thm2-alternative} 
The generating series of order $k\geq 3$ logarithmic derivatives of tau-function of a solution to the Toda lattice hierarchy has the expression 
\beq\label{alt-k-point}
\sum_{i_1, \dots, i_k=0}^\infty \frac1{\lambda_1^{i_1+2}\dots \lambda_k^{i_k+2}} \frac{\pal^k\log\tau_n({\bf t})}{\pal t_{i_1}\dots \pal t_{i_k}}
=-\sum_{\sigma\in S_{k-2}} \frac{\left \langle {R}_n({\bf t},\lambda_k) \,,\, {\rm ad}_{R_n({\bf t},\lambda_{\sigma_1})} \cdots {\rm ad}_{R_n({\bf t},\lambda_{\sigma_{k-2}})} {R}_n({\bf t},\lambda_{k-1})  \right\rangle}{ (\lambda_{\sigma_{k-2}}-\lambda_{k-1}) (\lambda_{k-1}-\lambda_{k}) (\lambda_{k}-\lambda_{\sigma_1})\prod_{j=1}^{k-3} (\lambda_{\sigma_j}-\lambda_{\sigma_{j+1}})}
\eeq 
where ${\rm ad}_a \, b := [a,b],$ and $\langle a,b\rangle := \tr \, a\,b.$
\end{cor}

\begin{remark}
The same type formula as \eqref{alt-k-point} holds true also for the generating series of \cite{BDY1,Zhou} for the Witten--Kontsevich correlators 
and for the correlators of Drinfeld--Sokolov hierarchies \cite{BDY2}.
\end{remark}

\setcounter{equation}{0}
\setcounter{theorem}{0}

\section{Computing GUE correlators} \label{s3}
In this section we prove Thm.\,\ref{thm3}, Thm.\,\ref{thm1} and present some examples. 

\setcounter{equation}{0}
\setcounter{theorem}{0}
\subsection{Proof of Thms.\,\ref{thm3} and \ref{thm1}}\par

We are to compute the matrix resolvent of the operator
$$
\Delta-\left( \begin{array}{cr}\lambda & -n\\ 1 & 0\end{array}\right).
$$
It follows from eq.\,\eqref{ham} that the formal series 
$$\gamma_n=\gamma(n,\lambda; {\bf t}=0)  = \sum_{j\geq 0} \frac{(2j-1)!! }{\lambda^{2j+1}}\, \gamma_{n,j}$$ 
satisfies
\beq\label{recur-gamma}
\gamma_{n+3}= 
 \frac{    \lambda ^2-(n+1)} {n+2} \, (\gamma_{n+2} -\gamma_{n+1})+ \frac{  n }{n+2 } \, \gamma_n
\eeq
along with the boundary conditions
\eqa
&& \gamma_{n,0} = 1, \quad \forall \, n\geq 0 \label{bound1} \\ 
&& \gamma_{0,j} = (-1)^j \, (2j-1)!!, \quad \forall\,j\geq 0. \label{bound2}
\eeqa
Clearly, solution to \eqref{recur-gamma}--\eqref{bound2} if exists must be unique.  
We are to show that 
$$
\gamma_n^*:= \sum_{j\geq 0} \frac{(2 j-1)!! }{\lambda^{2j+1}} \, _2F_1(-j,1-n,1;2), \qquad n\geq 0
$$
 satisfies \eqref{recur-gamma}--\eqref{bound2}. Indeed, from
\beq \label{gnj}
\gamma^*_{n,j}= \, _2F_1(-j,1-n;1;2) =  \sum_{i=0}^j 2^i \left( \begin{array}{c}j\\i\end{array}\right)\left(\begin{array}{c}n-1\\i\end{array}\right),\qquad \forall\, n,j\geq 0
\eeq
 it is easy to see that 
$$
\gamma^*_{n,0}=1,\qquad \gamma^*_{0,j}=(-1)^j \, (2j-1)!!.
$$
So eqs.\, \eqref{bound1}--\eqref{bound2} are verified.  Eq.\,\eqref{recur-gamma} is equivalent the following recursion on $\gamma_{n,j}$
$$
(n+2) \gamma_{n+3,j} = (2j+1) (\gamma_{n+2,j+1}-\gamma_{n+1,j+1}) -(n+1) (\gamma_{n+2,j}-\gamma_{n+1,j}) + n\, \gamma_{n,j}.
$$
To show $\gamma^*$ (see \eqref{gnj}) is a solution to the above equation, it suffices to show
\eqa
&& \!\!\!\!\!\! (n+2) \, _2F_1(-j,-n-2;1;2) = (2j+1) \left[ \, _2F_1(-j-1,-n-1;1;2)  -  \, _2F_1(-j-1,-n;1;2) \right] \nn\\ 
&& \!\!\!\!\!\!  \qquad \qquad - (n+1) \left[ \, _2F_1(-j,-n-1;1;2)  -\, _2F_1(-j,-n;1;2) \right] + n\,  _2F_1(-j,1-n;1;2). \nn\\ \label{remains}
\eeqa
We now use the following contiguous relations of Gauss:  $\forall \, a,b,c,z \in \mathbb{C}$
\eqa
&& \!\!\!\!\!\!\!\!\!\!\!\!\!\!\!\!\!\! (c-a) \, {}_2F_1(a-1,b;c;z) + (a-c+bz) {}_2F_1(a,b;c;z) \nn\\
&& \!\!\!\!\!\!\!\!\!\!\!\!\!\!\!\!\!\! \qquad\qquad\qquad =  (c-b) \, {}_2F_1(a,b-1;c;z)  +  (b-c+az) {}_2F_1(a,b;c;z) \label{C1} \\
&& \!\!\!\!\!\!\!\!\!\!\!\!\!\!\!\!\!\! a \,  \left[\,{}_2F_1(a+1,b;c;z) -  {}_2F_1(a,b;c;z)\right] = b \,  \left[\,{}_2F_1(a,b+1;c;z) -  {}_2F_1(a,b;c;z)\right]. \label{C2}
\eeqa
Taking in \eqref{C1} $c=1, \, a=-j, \, b=-n-1$ we obtain  
$$
(n+2) \, {}_2F_1(-j,-n-2;1;2) = (1+j) \, {}_2F_1(-j-1,-n-1;1;2) - (n+1-j) \, {}_2F_1(-j,-n-1;1;2).
$$
Taking  in \eqref{C1}  $c=1, \, a=-j, \, b=-n$ we obtain  
$$
(n+1) \, {}_2F_1(-j,-n-1;1;2) = (1+j) \, {}_2F_1(-j-1,-n;1;2) - (n-j) \, {}_2F_1(-j,-n;1;2).
$$
Taking  in \eqref{C1} $c=1, \, a=-j, \, b=-n+1$ we obtain  
\beq\label{C13}
n \, {}_2F_1(-j,-n;1;2) = (1+j) \, {}_2F_1(-j-1,1-n;1;2) - (n-1-j) \, {}_2F_1(-j,1-n;1;2).
\eeq
Taking in \eqref{C2} $c=1,\, a=-j-1, \, b=-n-1$ we obtain
$$
 (n-j)\, {}_2F_1(-j-1,-n-1;1;2) = (n+1) \,{}_2F_1(-j-1,-n;1;2) - (j+1) \,{}_2F_1(-j,-n-1;1;2).
$$
So we have 
\eqa
&&\mbox{l.h.s. of \eqref{remains}}\nn\\
&=&(1+j) \, {}_2F_1(-j-1,-n-1;1;2) - (n+1-j) \, {}_2F_1(-j,-n-1;1;2)\nn \\
&=& \frac{(1+j)(n+1)} {n-j} \, {}_2F_1(-j-1,-n;1;2) - \frac{ j^2+n+j+1+(n-j)^2} {n-j} \, {}_2F_1(-j,-n-1;1;2) \nn\\
&=& \frac{(1+j)(1+2j)} {n+1} \, {}_2F_1(-j-1,-n;1;2) + \frac{ j^2+n+j+1+(n-j)^2}{n+1}\, {}_2F_1(-j,-n;1;2), \nn
\eeqa
and we have
\eqa
&& \mbox{r.h.s. of \eqref{remains}} \nn\\
&=& \frac{(2j+1)(j+1)}{n-j} \, {}_2F_1(-j-1,-n;1;2) - \left(\frac{(2j+1)(j+1)}{n-j} +n+1\right)\, {}_2F_1(-j,-n-1;1;2)   \nn\\ 
&& \!\!\!\!\!\!  \qquad \qquad  +(n+1)\, _2F_1(-j,-n,1;2) + n\,  _2F_1(-j,1-n,1;2)\nn\\
&=& \frac{(2j-n)(j+1)}{n+1} \, {}_2F_1(-j-1,-n;1;2) + \left(\frac{(2j+1)(j+1)}{n+1} +2n-j+1\right)\, {}_2F_1(-j,-n;1;2)   \nn\\ 
&& \!\!\!\!\!\!  \qquad \qquad  + n\,  _2F_1(-j,1-n;1;2).\nn
\eeqa
Comparing the above equations and using \eqref{C13} we find it suffices to show
$$
{}_2F_1(-j-1,-n;1,2)= {}_2F_1(-j,-n;1,2) + {}_2F_1(-j-1,1-n;1;2) +{}_2F_1(-j,1-n;1;2)
$$
which can be verified easily. 

Finally, using \eqref{rr3} and considering $v_n=0,w_n=n$ we find
\beq\label{rec-alpha}
\lambda \, (\alpha_n-\alpha_{n+1}) = n\, \gamma^*_n - (n+1) \, \gamma^*_{n+2}.
\eeq
Similarly as above it can be verified that $\alpha^*$ defined by
$$
\alpha^*_n= n \sum_{j\geq 0} \frac{ (2j+1)!! }{ \lambda^{2j+2}} \, _2F_1(-j,1-n;2;2)
$$
is a solution to \eqref{rec-alpha}. Moreover, $\alpha^*$ obviously satisfies the boundary condition
$$
\alpha^*_0(\lambda)=0.
$$  
The theorem is proved.  \epf

\noindent \textit{Proof}. of Thm.\,\ref{thm1}. \quad The GUE partition function is a particular tau-function of the Toda lattice hierarchy; see 
Prop.\,\ref{partition-tau-GUE-toda} in Appendix \ref{appa} for a detailed proof.
The initial data of the corresponding solution satisfy $v_n=0,\,w_n=n$. As a result, Part 2) of the theorem readily follows 
from Theorems \ref{thm3} and \ref{thm2}. It remains to prove Part 1).  
By definition we have
\beq\label{rec-c-tau}
c_{n+1,j+1} =  \frac{\pal }{\pal t_{j} } \log \frac{\tau_{n+1}}{\tau_n} =  \frac{\pal }{\pal t_{j} } \log \tau_{n+1} -  \frac{\pal }{\pal t_{j} } \log \tau_{n},\qquad \forall\, j\geq 0.
\eeq
Taking ${\bf t}=0$ in \eqref{rec-c-tau} and recalling that 
$\gamma_n= \sum_{j\geq 0} \frac{(2 j-1)!! }{\lambda^{2j+1}} \, _2F_1(-j,1-n;1;2)$,
and using the boundary condition $C_1(0;\lambda)=0$ 
we obtain the expression \eqref{onepoint} for one-point correlators.

The theorem is proved. \epf

\setcounter{equation}{0}
\setcounter{theorem}{0}
\subsection{An algorithm for computing connected GUE correlators. Examples}\par
Based on Thm.\,\ref{thm1} and formula \eqref{naR}, we give in this subsection a recursive procedure of 
calculating connected GUE correlators, which is very efficient in computation. 
\begin{defi} \label{d-DR} Fix ${\bf b}=(b_1,b_2,b_3\dots)$ an arbitrary sequence of positive integers. 
Define recursively a family of Laurent series 
$R_{K}^{\bf b}(n,\lambda)\in {\rm Mat} \left(2,\mathbb{Z}[n]((\lambda^{-1}))\right)$ with $K=\{k_1,\dots,k_{m}\}$ by
\eqa
\!\!\!\!\! && R^{\bf b}_{\{\}}(n,\lambda):=\mathcal{R}_n(\lambda), \nn\\
\!\!\!\!\! && R^{\bf b}_{K}(n,\lambda) :=  \sum_{I\sqcup J=K - \{k_1\}}  \,  
\left[R^{\bf b}_{I}(n,\lambda), \left(\lambda^{b_{k_1}}\,R^{\bf b}_{J}(n,\lambda)\right)_+\right]. \label{int-rec}
\eeqa
Here $k_1,\dots,k_{m}$ are distinct positive integers, $m=|K|$, and $\mathcal{R}_n(\lambda)$ is defined by eq.\,\eqref{ram}.
\end{defi}
\begin{lemma} \label{l-DR}
In the particular case that $b_1=b_2=b_3=\dots=b$ we have
$$
R^{\bf b}_{K}(n,\lambda)=R^{\bf b}_{K'}(n,\lambda)=:R^b_{|K|}(n,\lambda), \qquad \mbox{as long as ~} |K|=|K'|. 
$$
Moreover, the following formulae hold true for $R^b_m(n,\lambda),\,m\geq 1$
\beq R^b_m(n,\lambda) =  \sum_{i=0}^{m-1}  
 \, \left( \begin{array}{c} m-1 \\ i \\ \end{array} \right) 
\left[R^b_{i}(n,\lambda), \left(\lambda^b\, R^b_{m-1-i}(n,\lambda)\right)_+\right].\nn
\eeq
\end{lemma}
Clearly, $R^b_0(n,\lambda)=\mathcal{R}_n(\lambda)$.
\begin{prop} \label{p-DR} Let ${\bf b}=(b_1,b_2,b_3\dots)$ be any sequence of positive integers, and $K=\{k_1,\dots,k_m\}$ any finite set of positive integers.
The following formula holds true for connected GUE correlators
\beq
\sum_{i,j\geq 1} \frac{\langle \tr \,M^{b_{k_1}} \cdots \tr \,M^{b_{k_{m}}} \, \tr \,M^i\, \tr \,M^j\rangle_c}{\lambda_1^{i+2}\lambda_2^{j+2}}  =  \sum_{I \sqcup J=K} \,
\, \frac{\tr \, R_{I}^{{\bf b}}(N,\lambda_1)\, R_{J}^{{\bf b}}(N,\lambda_2)}{(\lambda_1-\lambda_2)^2} - \frac{\delta_{m,0} }{(\lambda_1-\lambda_2)^2}.
\eeq
Here $m=|K|$. In the particular case that $b_1=b_2=\dots=b$ for some $b\geq 1$, we have $\forall\,m\geq 0$
\beq\label{ij-bk}
\sum_{i,j\geq 1} \frac{\left\langle \left(\tr \, M^{b}\right)^m \, \tr \, M^i\, \tr \, M^j \right\rangle_c}{\lambda_1^{i+1}\lambda_2^{j+1}}  =  \sum_{i=0}^m \,\left( \begin{array}{c} m \\ i \\ \end{array} \right) 
\, \frac{\tr \, R^b_{i}(N,\lambda_1)\, R^b_{m-i}(N,\lambda_2)}{(\lambda_1-\lambda_2)^2} - \frac{\delta_{m,0} }{(\lambda_1-\lambda_2)^2}.
\eeq
\end{prop}
\noindent \textit{Proof}. By using mathematical induction and by noticing that the term containing $Q_n(\mu)$ in r.h.s. of \eqref{naR} does not contribute to
generating series of correlators as it was proved in Thm.\,\ref{thm2}.  \epf
Clearly, Def.\,\ref{d-DR}, Lem.\,\ref{l-DR} and Prop.\,\ref{p-DR} give an algorithm of computing connected GUE correlators.
Similar recursive formulation as \eqref{int-rec}--\eqref{ij-bk} also works for logarithmic derivatives of tau-function  of an arbitrary solution to the Toda lattice hierarchy.

\begin{primer}  [$1$-point correlators] We have $\left\langle \tr \, M^{a+1} \right \rangle_c=0$
if $a$ is an even integer then  vanishes; otherwise,
\beq \label{onepoint-cor}
\left\langle  \tr\, M^{a+1} \right \rangle_c = N \, a!!  \left[\, _2F_1\left(-\frac{a+1}2,- N; 2;2\right) -  \frac{a+1}2  \, _2F_1\left(-\frac{a-1}2,1-N;3;2\right)\right],\quad a\geq 1.
\eeq
For example,
$$
\left\langle  \tr\, M^{2} \right \rangle_c = N^2, \qquad \left\langle  \tr\, M^{4} \right \rangle_c = N+2N^3, \qquad  \left\langle  \tr\, M^{6} \right \rangle_c = 10 N^2+ 5 N^4,
$$
$$
\left\langle  \tr\, M^{20} \right \rangle_c = 16796 N^{11}+1385670 N^9+31039008 N^7+211083730 N^5+351683046 N^3+59520825 N.
$$
\end{primer}

\begin{primer} [$2$-point correlators] $\left\langle  \tr\, M^{a+1}\, \tr\, M^{b+1} \right\rangle_c$ vanishes
 $\forall \, a,b\geq 0,$ if  $a+b $ is an odd integer; otherwise,
\eqa
&& \!\!\!\!\!\! ~ \left\langle  \tr\, M^{a+1} \, \tr\, M^{b+1} \right\rangle_c \nn\\
&& \!\!\!\!\!\!\!\! =  N \, (a+b+1)!! \, (1+b) \, _2F_1\left(-\frac{a+b}2, 1-N;2;2\right) \nn\\
&& \!\!\!\!\!\!\!\! +  2 N^2\, \sum_{0\leq j\leq b-2 \atop j \equiv b ( {\rm mod} 2)} (a+j+1)!!  (b-j-1)!! \, (1+j)  \,
 _2F_1\left(-\frac{a+j}2,1-N;2;2\right) \, _2F_1 \left(-\frac{b-j-2}{2},1-N;2;2\right) \nn\\
 &&\!\!\!\!\!\!\!\!  - N  \sum_{0\leq j\leq b-1 \atop j \equiv b-1 ({\rm mod} 2)} (a+j)!!(b-j-2)!!\, (1+j)\,  \bigg[ {}_2F_1\left(-\frac{a+1+j}2,-N;1;2\right) \, _2F_1\left(-\frac{b-j-1}2,1-N;1;2\right) \nn\\
&& \qquad\qquad\qquad\qquad + _2F_1\left(-\frac{b-j-1}2,-N;1;2\right) \, _2F_1\left(-\frac{a+1+j}2,1-N;1;2\right) \bigg].  \nn\\  \label{two-point-cor}
\eeqa
For example, we have
$$
\left\langle  \tr\, M\, \tr\, M \right \rangle_c = N, \qquad \left\langle  \tr\, M^{2}\, \tr\, M^4 \right \rangle_c =4N + 8N^3, \qquad  \left\langle  \tr\, M^{4}\, \tr\, M^4 \right \rangle_c =60 N^2+ 36 N^4,
$$
\eqa
&&\!\!\!\!\! \left\langle  \tr\, M^{18}\, \tr\, M^{20} \right \rangle_c \nn\\
&&\!\!\!\!\! = 4813380 N\, (8840 N^{18}+3275220 N^{16}+478887552 N^{14}+34305326120 N^{12}+1259109855744 N^{10}\nn\\
&&\!\!\!\!\! +23197400694000 N^8+199375600144496 N^6+689468897044260 N^4+705221681016618 N^2\nn\\
&&\!\!\!\!\! +85187495274525). \nn
\eeqa
\end{primer}

\begin{primer} [$3$-point correlators] We have $\forall\, i\geq 1,$
\eqa
&& \!\!\!\!\!  \left\langle\left(\tr\, M^2\right)^2  \tr \, M^i\right\rangle_c=N\, {\rm Coef}_{\lambda^{-i-1}}   \left[
\left(\lambda^2-1\right)  \gamma_{N+1}(\lambda)-\left(\lambda^2+1\right) N \, \gamma_N(\lambda)\right]\nn\\
&& \!\!\!\!\!  \left\langle\tr\, M^2 \, \tr\,M^3 \, \tr \, M^i \right\rangle_c= N \, {\rm Coef}_{\lambda^{-i-1}}   \left[ \,
4 \, \alpha_N(\lambda)-\lambda \left(\lambda^2+2 N+2\right) \gamma_N(\lambda)+\lambda \left(\lambda^2+2 N -2\right) \gamma_{N+1}(\lambda)+2\right]\nn\\
&& \!\!\!\!\!  \left\langle\left(\tr\, M^3\right)^2  \tr \, M^i\right\rangle_c = N\, {\rm Coef}_{\lambda^{-i-1}}   
 \left[ \, 8 \, \lambda \, \alpha_N(\lambda)-\left(\lambda^4+ 4 N^2+\lambda^2 (4 N+3)+8 N+3\right) \gamma_N(\lambda) \right. \nn\\
&& \!\!\!\!\!   \qquad\qquad\qquad\qquad\qquad \left.+\left(\lambda^4+4 N^2+\lambda^2 (4 N - 3)-8 N+3\right)
\gamma_{N+1}(\lambda)+4 \lambda\right]\nn\\
&& \!\!\!\!\!  \left\langle\left(\tr\, M^4\right)^2 \tr \, M^i\right\rangle_c= N\,{\rm Coef}_{\lambda^{-i-1}} \left[ \,4 \lambda \left(\lambda^2+6 N\right) (2 \alpha_N(\lambda)+1) \right.\nn\\
&& \!\!\!\!\! \qquad\qquad \left. -\left(\lambda^6+\lambda^4 (4 N+3)+\lambda^2 (2 N+1) (2 N+9)+36 N (N+1)+15\right) \gamma_N(\lambda)\right.\nn\\
&&  \!\!\!\!\! \qquad\qquad \left.+\left(\lambda^6+\lambda^4 (4 N-3)+\lambda^2 (2N-1)(2N-9)-36 (N-1) N-15\right) \gamma_{N+1}(\lambda)\right]. \nn
\eeqa
Here we recall  
$$
\alpha_n(\lambda)= n \sum_{j\geq 0} \frac{ (2j+1)!! }{ \lambda^{2j+2}} \, _2F_1(-j,1-n;2;2), \quad 
\gamma_n(\lambda)= \sum_{j\geq 0} \frac{(2 j-1)!! }{\lambda^{2j+1}} \, _2F_1(-j,1-n,1;2). 
$$
For example,
$$
\left\langle \left( \tr\, M^2\right)^3 \right\rangle_c = 8N^2, \qquad  \left\langle \left( \tr\, M^2\right)^2 \tr\, M^4 \right\rangle_c = 24N+48N^3,
$$
$$
\left\langle  \left(\tr\, M^2\right)^2 \tr\, M^6 \right\rangle_c = 480N^2+240N^4, \qquad  \left\langle  \left(\tr\, M^2\right)^2 \tr\, M^8 \right\rangle_c = 1680N+5600N^3+1120N^5,
$$
\eqa
\!\!\!\!\! \left\langle \left(\tr \, M^2\right)^2 \tr \, M^{38}\right\rangle_c &=& 1343120024400 \, N^2 \left(2 N^{18}+1140 N^{16}+240312 N^{14}+24082880 N^{12}\right.\nn\\
&&\!\!\!\!\! \left.+1231558302 N^{10}+32196168420 N^8+410364369452 N^6\right.\nn\\
&&\!\!\!\!\! \left.+2294179050960 N^4+4562960651307 N^2+1979828515350\right).\nn
\eeqa
$$
\left\langle  \tr\, M^2\, \left(\tr\, M^3\right)^2 \right\rangle_c =18N+72N^3 , \quad \langle  \tr\, M^2\, \tr\, M^3\, \tr\, M^5 \rangle_c = 480N^2+360 N^4, 
$$
$$
\langle  \tr\, M^2\, \tr\, M^3\, \tr\, M^7 \rangle_c=480N^2+360N^4, \quad \langle  \tr\, M^2\, \tr\, M^3\, \tr\, M^9 \rangle_c = 1470 N+6300N^3+1680 N^5, 
$$
\eqa
\!\!\!\!\!\langle  \tr\, M^2\, \tr\, M^3\, \tr\, M^{39} \rangle_c&=&1033848966150 N \left(16 N^{20}+10108 N^{18}+2401182 N^{16}+276911776 N^{14}\right.\nn\\
&&\!\!\!\!\! \left.+16743310948 N^{12}+536717003004 N^{10}+8831088179794 N^8\right.\nn\\
&&\!\!\!\!\! \left.+68958855149632 N^6+219890931285060 N^4+210352383917730 N^2\right.\nn\\
&& \!\!\!\!\! \left.+24130040059125\right).\nn
\eeqa
$$
\left\langle  \left(\tr\, M^3\right)^2 \tr\, M^4 \right\rangle_c = 468N^2+432 N^4,  \quad \left\langle  \left(\tr\, M^3\right)^2 \tr\, M^6 \right\rangle_c = 1350N+6660N^3+2160 N^5,
$$
$$
\left\langle  \left(\tr\, M^3\right)^2 \tr\, M^8 \right\rangle_c = 55440 N^2+ 68040N^4+10080 N^6,$$
$$
\left\langle  \left(\tr\, M^3\right)^2 \tr\, M^{10} \right\rangle_c = 213570N+1183140 N^3+570780 N^5+45360 N^7,
$$
\eqa
\!\!\!\!\! \left\langle \left( \tr\, M^3\right)^2 \tr\, M^{38} \right\rangle_c &=& 1511010027450 N \left(16 N^{20}+9628 N^{18}+2180250 N^{16}+239934736 N^{14}\right.\nn\\
&& \!\!\!\!\! \left. +13863233644 N^{12}+425408903244 N^{10}+6715474080598 N^8\right.\nn\\
&& \!\!\!\!\! \left. +50449385602192 N^6+155303372658492 N^4+144060538320450 N^2\right.\nn\\
&& \!\!\!\!\! \left. +16119257529375\right).\nn
\eeqa
$$\left\langle  \tr\, M^2\, \left(\tr\, M^4\right)^2  \right\rangle_c = 480 N^2 + 288 N^4,  \quad \left\langle  \left(\tr\, M^4\right)^3 \right\rangle_c = 1728 N^5+6336 N^3+1440 N,$$
$$\left\langle  \left(\tr\, M^4\right)^2 \tr\, M^6 \right\rangle_c = 8640 N^6+63360 N^4+56160 N^2,$$
$$\left\langle  \left(\tr\, M^4\right)^2 \tr\, M^{8} \right\rangle_c = 40320 N^7+530880 N^5+1162560 N^3+221760 N,$$
\eqa
\!\!\!\!\! \left\langle  \left(\tr\, M^4\right)^2 \tr\, M^{38} \right\rangle_c &=& 8058720146400 N^2 \left(12 N^{20}+8408 N^{18}+2249790 N^{16}+297878352 N^{14}\right.\nn\\
&& \!\!\!\!\! \left. +21183159128 N^{12}+824144717136 N^{10}+17179527894426 N^8\right.\nn\\
&& \!\!\!\!\! \left. +180912770249240 N^6+860693336297694 N^4+1496297650892364 N^2\right.\nn\\
&& \!\!\!\!\! \left. +582832451267325\right). \nn
\eeqa
\end{primer}
\begin{primer} [$4$-point correlators] We have $\forall\,i\geq 1$,
\eqa
 \!\!\!\!\! && \left\langle \left(\tr\, M^2\right)^3  \tr \, M^i\right\rangle_c  =\,  N\, {\rm Coef}_{\lambda^{-i-1}}  \left[ -4 \lambda^3 \alpha_N(\lambda)+\left(\lambda^4-3\right) (\gamma_N(\lambda)+ \gamma_{N+1}(\lambda) )-2 \lambda^3\right]\nn\\
 \!\!\!\!\! && \left\langle \left(\tr\, M^3\right)^3  \tr \, M^i\right\rangle_c  =\,  N\, {\rm Coef}_{\lambda^{-i-1}} \nn\\
 && \!\!\!\!\!  \left[-2 \, \left(\lambda^6-3 \lambda^2+8 N^3+12 \lambda^2 N^2+6 \lambda^4 N-38 N\right) \, (2\,\alpha_N(\lambda) +1)\right.\nn\\
 && \!\!\!\!\! \left. +  \left(\lambda^7+3 \lambda^3 \left(4 N^2-2 N-1\right)+3 \lambda^5 (2 N+1)+2\lambda (N-5) (2 N+1) (2 N+3)\right) \gamma_N(\lambda) \right. \nn\\
 && \!\!\!\!\!  \left.+  \left(\lambda^7+3 \lambda^3 \left(4 N^2+2N-1\right)+3 \lambda^5 \, (2 N-1)+2\lambda (N+5)(2N-1)(2N-3)  \right) \gamma_{N+1}(\lambda) \right] \nn\\
 \!\!\!\!\! && \left\langle \left(\tr\, M^4\right)^3  \tr \, M^i\right\rangle_c  =\,  N\, {\rm Coef}_{\lambda^{-i-1}} \nn\\
 && \!\!\!\!\!  \left[  \, \alpha_N(\lambda) \, \left( \lambda  \, (828+2160N^2) + \lambda^3 (152N-32N^3) -24 \lambda^5(1+2N^2)-24 N \lambda^7  -4 \lambda^9 \right)\right.\nn\\
 && \!\!\!\!\! \left. + \, \gamma_N(\lambda)\, \left(\lambda^{10}+3 \lambda^8(1+2N)+6 \lambda^6(1+2N+2N^2)-2 \lambda^4(15+30N^2-4N^3)\right.\right.\nn\\
  && \!\!\!\!\! \left. \left. -(315+756N+828N^2+144N^3) \lambda^2-1944 N^3 -2916 N^2-56 \lambda^4 N-2862 N-945\right)  \right. \nn\\
 && \!\!\!\!\!  \left.+\, \gamma_{N+1}(\lambda)\, \left(\lambda^{10}-3 \lambda^8(1-2N)+6 \lambda^6(1-2N+2N^2)+2 \lambda^4(15+30N^2+4N^3)\right.\right. \nn\\
&& \!\!\!\!\!  \left. \left. -(315-756N+828N^2-144N^3) \lambda^2-1944 N^3+2916 N^2-56 \lambda^4 N-2862 N+945\right) \right.\nn\\
&& \!\!\!\!\!  \left. -2 \lambda^9 -12 \lambda^7 N -12(1+2N^2) \lambda^5 + (76N-16N^3) \lambda^3 + (414+1080N^2) \lambda  \right]. \nn
\eeqa

For example,
$$\left\langle\left(\tr\, M^2\right)^4\right\rangle_c = 48N^2, \qquad \left\langle\left(\tr\, M^2\right)^3 \, \tr \, M^4  \right\rangle_c = 192N+384N^3,$$
$$\left\langle\left(\tr\, M^2\right)^3\, \tr\, \, M^{20}\right\rangle_c =44341440 N \left(4 N^{10}+330 N^8+7392 N^6+50270 N^4+83754 N^2+14175\right),$$
$$\left\langle\left(\tr\, M^3\right)^4\right\rangle_c = 4536 N^2 + 5184 N^4, \qquad \left\langle\left(\tr\, M^3\right)^3 \, \tr \, M^5  \right\rangle_c =15390 N + 82620 N^3 + 32400 N^5,$$
$$
\left\langle\left(\tr\, M^3\right)^3 \, \tr \, M^{17}  \right\rangle_c=149652360 N^2 \left(16 N^{10}+1354 N^8+33462 N^6+282518 N^4+730832 N^2+374043\right),
$$
$$\left\langle\left(\tr\, M^4\right)^3 \, \tr\,M^2\right\rangle_c =17280N+76032 N^3+20736 N^5 , \qquad \left\langle\left(\tr\, M^4\right)^4  \right\rangle_c =770688N^2+964224N^4+145152N^6,$$
\eqa
 \!\!\!\!\! \left\langle\left(\tr\, M^4\right)^3 \, \tr \, M^{20}  \right\rangle_c&=&798145920 N^2 \left(60 N^{12}+8674 N^{10}+402650 N^8+7343262 N^6+51873380 N^4\right.\nn\\
&& \!\!\!\!\!  \left.+120454639 N^2+57830535\right). \nn
\eeqa
\end{primer}

\begin{primer} [$5$-point correlators] We have $\forall\,i\geq 1$,
\eqa
 \!\!\!\!\! && \left\langle \left(\tr\, M^2\right)^4  \tr \, M^i\right\rangle_c  =\,  N\, {\rm Coef}_{\lambda^{-i-1}} \nn\\
 && \!\!\!\!\!  \left[\left(\lambda^6+\lambda^4-3 \lambda^2-4 \lambda^4 N-15\right) \gamma_{N+1}(\lambda)- \left(\lambda^6-\lambda^4-3 \lambda^2-4 \lambda^4 N+15\right) \gamma_N(\lambda)-8 \lambda^3 \alpha_N(\lambda)-4 \lambda^3 \right]\nn\\
 \!\!\!\!\! && \left\langle \left(\tr\, M^3\right)^4  \tr \, M^i\right\rangle_c  =\,  N\, {\rm Coef}_{\lambda^{-i-1}} \nn\\
 && \!\!\!\!\!  \left[8 \lambda \left(\lambda^6-64 N^3-3 \lambda^2 \left(20 N^2+7\right)-12 \lambda^4 N+304 N\right) \alpha_N(\lambda) \right. \nn\\
 && \!\!\!\!\! \left. +\left(\lambda^{10}+315 \left(\lambda^2-3\right)-4 N \left(2 N \left(2 N \left(4 N^2+N-106\right)+463\right)-687\right)+\lambda^8 (4 N-3) \right.\right.\nn\\
  && \!\!\!\!\! \left. \left. -\lambda^6 (8 (N-2) N+9)+\lambda^4 (45-8 N (4 N (2 N-3)+13))+8 \lambda^2 N (2 N (N (6-7 N)+49)-93)\right) \gamma_{N+1}(\lambda) \right.\nn\\
  && \!\!\!\!\! \left. +\left(-\lambda^{10}-315 \left(\lambda^2+3\right)-\lambda^8 (4 N+3)+\lambda^6 (8 N (N+2)+9)+\lambda^4 (8 N (4 N (2 N+3)+13)+45) \right. \right. \nn\\
    && \!\!\!\!\! \left. \left.+8 \lambda^2 N (2 N (N (7 N+6)-49)-93)+4 N (2 N (2 N (N (4 N-1)-106)-463)-687)\right) \gamma_N(\lambda) \right.\nn\\
      && \!\!\!\!\! \left. +4 \lambda \left(\lambda^6-64 N^3-3 \lambda^2 \left(20 N^2+7\right)-12 \lambda^4 N+304 N\right)\right]. \nn
\eeqa
For example, 
$$\left\langle\left(\tr\, M^2\right)^5\right\rangle_c = 384 N^2, \qquad \left\langle\left(\tr\, M^2\right)^4 \, \tr \, M^4  \right\rangle_c = 1920 N + 3840 N^3,$$
$$\left\langle\left(\tr\, M^2\right)^4\, \tr\, \, M^{20}\right\rangle_c =1152877440 N \left(4 N^{10}+330 N^8+7392 N^6+50270 N^4+83754 N^2+14175\right),$$
$$\left\langle\left(\tr\, M^3\right)^4\, \tr\, \, M^{2} \right\rangle_c = 62208 N^4+54432 N^2, \qquad \left\langle\left(\tr\, M^3\right)^4 \, \tr \, M^4  \right\rangle_c =528768 N^5+1181952 N^3+204120 N,$$
$$
\left\langle\left(\tr\, M^3\right)^4 \, \tr \, M^{18}  \right\rangle_c = 283551840 N^2 (5440827 + 11132606 N^2 + 4554930 N^4 + 576009 N^6 + 
   25058 N^8 + 320 N^{10}).
$$
\end{primer}

\medskip

The GUE correlators $\langle\tr\, M^{i_1}\, \dots \tr\, M^{i_k}\rangle_c$ that we compute 
 with $i_1+\dots+i_k\leq 10$ agree with \cite{AMM,BIZ,MS}. 

\subsection{Polygon numbers on Riemann surfaces}
In this subsection, we consider correlators of the form
$$
\left\langle \left(\tr \, M^b\right)^k \right\rangle_c,\qquad b\geq 3, \, k\geq 1
$$
which is exactly the case when $b_1=b_2=\dots=b$ in the algorithm described by Def.\,\ref{d-DR}--Prop.\,\ref{p-DR}.

These correlators are polynomials in $N$ whose coefficients are positive integers, called \textit{polygon numbers} on Riemann surfaces.
More precisely, we have
$$\left\langle  \left(\tr \, M^{b}\right)^k \right \rangle_c\,=\sum_{0\leq g \leq \frac{k}4(b-2)+\frac12} \, n_{g,b,k} \, N^{2-2g + (\frac b2-1) k}$$
where $n_{g,b,k}$ counts the number of connected oriented labelled ribbon graphs of genus $g$ with $k$ vertices of valencies $b.$
In other words, $n_{g,b,k}$ counts the number of labelled \textit{maps} which are embedded into a connected oriented closed surface of genus $g$; see details for example in \cite{BIZ,EM}, or  Appendix A.

Using our algorithm we worked out a program computing the polygon numbers. Below are tables of first several polygon numbers with 
$b=3,4,5,6,7,8$.

\begin{table}[!htbp]
\begin{center}\tiny 
    \begin{tabular} {|c|R{2.0cm}|R{2.0cm}|R{2.0cm}|R{2.0cm}|R{2.0cm}|R{2.0cm}|N}
    \hline
    $k$ & $g=0$ & $g=1$ & $g=2$ &$g=3$ & $g=4$ & $g=5$ & \\[8pt]
    \hline
    $2$ & $12$ & $3$& $0$     &$0$      & $0$ & $0$  & \\[8pt]
    \hline
    $4$ & $5184$ & $4536$     & $0$  & $0$ & $0$ & $0$ & \\[8pt]
   \hline
    $6$ & $9797760$ & $19362240$ & $3061800$          & $0$  & $0$ & $0$ & \\[8pt]
    \hline
    $8$ & $45148078080$ & $164367221760$ & $89414357760$  & $0$ & $0$  &  $0$ & \\[8pt]
    \hline
   $10$  & $392212641300480$  & $2332019568291840$ &  $2834113460935680$ & $357485480352000$  & $0$ & $0$ & \\[8pt]
    \hline
    $12$ & 5 560 971 849 577 267 200 & 49 838 762 032 083 763 200 & 110 757 832 882 937 856 000 & 47 537 982 337 808 793 600 &  0 & 0 & \\[16pt]
    \hline
    \end{tabular}
\end{center}
\caption{Triangle numbers $n_{g,3,k}$.} \label{numbers-3}
\end{table}
The above numbers of triangulations agree with the ones computed by Fleming \cite{fle}, and with Table \ref{numbers-DZ} in Appendix A.

\begin{table}[!htbp]
\begin{center}\tiny 
    \begin{tabular} {|c|R{2.0cm}|R{2.0cm}|R{2.0cm}|R{2.0cm}|R{2.0cm}|R{2.0cm}|N}
    \hline
     $k$ & $g=0$ & $g=1$ & $g=2$ &$g=3$ & $g=4$ & $g=5$ & \\[8pt]
    \hline
    $1$ & $2$ & $1$ & $0$   & $0$ & $0$    & $0$ & \\[8pt]
    \hline
    $2$ & $36$ & $60$& $0$     &$0$      & $0$ & $0$  & \\[8pt]
    \hline
    $3$ & $1728$  & $6336$  & $1440$  &$0$     			             & $0$ & $0$ & \\[8pt]
    \hline
    $4$ & $145152$ & $964224$     & $770688$  & $0$ & $0$ & $0$ & \\[8pt]
    \hline
    $5$ & $17915904$ & $192098304$   & $348033024$ & $58060800$ & $0$ & $0$ & \\[8pt]
    \hline
    $6$ & $2956124160$ & $47357706240$ & $158525890560$     & $92253634560$  & $0$ & $0$ & \\[8pt]
    \hline
    $7$ & $614873825280 $ & $13922807316480$  
    & $76300251955200$ & $100275872071680$ & $13948526592000$ & $0$ & \\[8pt]
    \hline
    $8$ & $154928203970560$ & $4755537360322$ $560$ & $39364669475389$ $440$  & $95431198231756$ $800$ & $45881115652915$ $200$  &  0 & \\[16pt]
    \hline
    $9$  & 45 977 357 978 173 440  & 1 850 918 058 999 152 640 &  21 844 654 140 570 992 640 & 86 654 328 700 277 882 880  & 93 561 769 862 061 096 960 
    & 11 473 053 680 664 576 000 & \\[16pt]
    \hline
    \end{tabular}
\end{center}
\caption{Quadrangle numbers $n_{g,4,k}$.} \label{numbers-4}
\end{table}
The numbers of quadrangulations (Table \ref{numbers-4}) for $k\leq 6$ agree with the ones computed by Pierce \cite{Pierce}.

\begin{table}[!htbp]
\begin{center}\tiny 
    \begin{tabular} {|c|R{2.0cm}|R{2.0cm}|R{2.0cm}|R{2.0cm}|R{2.0cm}|R{2.0cm}|N}
    \hline
     $k$ & $g=0$ & $g=1$ & $g=2$ &$g=3$ & $g=4$ & $g=5$ & \\[8pt]
     \hline
    $2$ & 180 & 600 & 165     &$0$      & $0$ & $0$  & \\[8pt]  \hline
    $4$ & $6480000$ & $93960000$     & $332100000$  & $219510000$ & $0$ & $0$ & \\[8pt] \hline
    $6$ & $1242216000000$ & $45300060000000$ & $546671268000000$          & $2354983470000000$  & $2843338018500000$ & $389492853750000$ & \\[8pt]
    \hline
    $8$ & $6246072$ $00000000000$ & $4472187552$ $0000000000$ & $12283912451664$ $00000000$  & $152254618488$ $00000000000$ & 
    $814486310134308$ $00000000$  &  $155872936216116$ $000000000$ & \\[16pt]
    \hline
    $10$ & 6135286523184 00000000000 & 748085330014848 00000000000 & 381919155214554 7200000000000  & 994894820469295 20000000000000 & 
    1345274552969624 982600000000000  &  886998667042254 5388000000000000 & \\[16pt]
    \hline
    \end{tabular}
\end{center}
\caption{Pentagon numbers $n_{g,5,k}$.} \label{numbers-5}
\end{table}

\begin{table}[!htbp]
\begin{center}\tiny 
    \begin{tabular} {|c|R{2.0cm}|R{2.0cm}|R{2.0cm}|R{2.0cm}|R{2.0cm}|R{2.0cm}|N}
    \hline
    $k$ & $g=0$ & $g=1$ & $g=2$ &$g=3$ & $g=4$ & $g=5$ & \\[8pt]  \hline
    1 & 5 & 10 & 0& 0& 0& 0 &  \\[8pt]\hline
    2 & 600 & 4800 & 4770     &$0$      & $0$ & $0$  & \\[8pt]  \hline
    3 & $216000$ & $4176000$     & $17290800$  & $12315600$ & $0$ & $0$ & \\[8pt] \hline
    4 & $142560000$ & $5287680000$ & $54015984000$          & $161062992000$  & $93360956400$ & $0$ & \\[8pt]
    \hline
    $5$ & $141523200000$  & $8805542400000$  & $174855024000000$  & $1291104489600000$ & 
    $3123016385040000$  &  $1565262377280000$ & \\[8pt]
    \hline
    $6$ & $190356480000000$ & $1819210752$ $0000000$ & $611671917312$ $000000$  & $8806826030976$ $000000$ & 
    $527219331093504$ $00000$  &  $1096721661511872$ $00000$ & \\[16pt]
     \hline
    $7$ & $3255095808$ $00000000$  & $448921046016$ $00000000$  & $233492421222144$ $0000000$  & $5707503816562944$ $0000000$ & 
    $666456378352813$ $440000000$  &  $34239929870156$ $77440000000$ & \\[16pt]
     \hline
    \end{tabular}
\end{center}
\caption{Hexagon numbers $n_{g,6,k}$.} \label{numbers-6}
\end{table}

\begin{table}[!htbp]
\begin{center}\tiny 
    \begin{tabular} {|c|R{2.0cm}|R{2.0cm}|R{2.0cm}|R{2.0cm}|R{2.0cm}|R{2.0cm}|N}
    \hline
     $k$ & $g=0$ & $g=1$ & $g=2$ &$g=3$ & $g=4$ & $g=5$ & \\[8pt]
     \hline
     $2$ & $2800$ & $34300$& $81340$     &$16695$      & $0$ & $0$  & \\[8pt]  \hline  
     $4$ & $4609920000$ & $270256560000$ & $5470015824000$      & $42516370176000$  & $108544213999200$ & $56597795793000$ & \\[8pt]    \hline
     $6$ & $43505659008$ $000000$  & $66630746448$ $00000000$  & $4225455365922$ $00000000$  & $134784066695700$ $244000000$ & 
     $219133289516560$ $146000000$  &  16984808089605 44078400000  & \\[16pt] \hline
     $8$ & 11023506707447 80800000000 & 3382196457804530 68800000000 & 4700014305292697 9005440000000 & 369200898096647 5731179520000000 & 17291108386034951 6876140800000000 & 48125851798487907 98122421760000000 & \\[16pt]
     \hline
    \end{tabular}
\end{center}
\caption{Heptagon numbers $n_{g,7,k}$.} \label{numbers-7}
\end{table}

\begin{table}[!htbp]
\begin{center}\tiny 
    \begin{tabular} {|c|R{2.0cm}|R{2.0cm}|R{2.0cm}|R{2.0cm}|R{2.1cm}|R{2.1cm}|N}
    \hline
     $k$ & $g=0$ & $g=1$ & $g=2$ &$g=3$ & $g=4$ & $g=5$ & \\[8pt] \hline
    1 & 14 & 70 & 21 & 0 & 0 & 0 &  \\[8pt]\hline
    $2$ & $9800$ & $215600$& $1009400$     & $781200$      & $0$ & $0$  & \\[8pt]  \hline
    $3$ & $21952000$ & $1218336000$     & $20217792000$  & $110898368000$ & $158932166400$ & $24309331200$ & \\[8pt] \hline
    $4$ & $92198400000$ & $10058845440000$ & $386873706240000$      & $6319266481920000$  & $42291774083328000$ & $96422698084608000$ & \\[8pt]
    \hline
    $5$ & $588594585600000$  & $1094166355968$ $00000$  & $79095340412928$ $00000$  & $277053418672128$ $000000$ & 
    $482663835053568$ $0000000$  &  $39425239788834$ $816000000$ & \\[16pt]
    \hline
    \end{tabular}
\end{center}
\caption{Octagon numbers $n_{g,8,k}$.} \label{numbers-8}
\end{table}

Most of the numbers in Tables \ref{numbers-5}--\ref{numbers-8} seem not to be computed in the literature.

\newpage

\section{Calculating the GUE correlators from the two-dimensional Frobenius manifold of the Toda lattice} \label{s4}
In this section, we briefly outline following \cite{DZ, Du1, icmp} 
 an algorithm for computing the genus expansion of the GUE free energy. 
It is just a specification of the general algorithm of \cite{DZ, Du1} applied to the two-dimensional Frobenius manifold with the potential 
$$
F=\frac12 u\, v^2 + e^u.
$$
Such a Frobenius manifold appears in the description of the structure of the long wave limit of the Toda lattice. 
The algorithm of \cite{DZ} proves to be quite powerful for computation of the low genera multipoint correlators. We have used it for checking the explicit examples presented above. 

The algorithm will be illustrated on computation of the weighted numbers of triangulations on surfaces of genus 0, 1 and 2. We will only describe the main steps of the algorithm referring the reader to \cite{DZ} for details.

Let $v=v(x,s)$ be the solution to the cubic equation 
\beq\label{cubic}
v\left( 1-9 s\, v + 18 s^2 v^2\right) =6 s\, x
\eeq
in the form of a power series in $s$ vanishing at $s=0$,
\beq\label{vv}
v=6\,s\, x + 324 \,s^3 x^2 +31104 \,s^5 x^3+\dots.
\eeq
Put
\beq\label{ww}
w=\frac{x}{1-6 \,s\, v} =x\left( 1+36\, s^2 x + 3240\, s^4 x^2+\dots\right).
\eeq
Introduce also the series
\beq\label{log}
u=\log w=\log x + 36 \, s^2 x +2592\, s^4 x^2 +\dots.
\eeq
Then the genus zero free energy is given by the series expansion of the following expression
\eqa\label{f0}
\mathcal{F}_0&=&\frac12 \left(v^2 w + \frac12 w^2\right) - 6 s \left(\frac12 v^3 w + v w^2\right) + 
 18\, s^2 \left(\frac14 v^4 w + v^2 w^2 + \frac13 w^3\right)
 \nn\\
 &&
 - x\, \left(\frac12 v^2 + w\right)  + 
 6\, s\, x \left(\frac16 v^3 + v \,w\right)  + \frac12 u \, x^2
 \\
 &=& \frac12 x^2 \left( \log x -\frac32\right)+6\, s^2 x^3 + 216\, s^4 x^4 +\dots.
 \nn
 \eeqa
Recall that the coefficient of $s^k$ at $x=1$ is equal to the weighted number of planar triangulations with $k$ triangles.

The genus one free energy is given by the formula
\eqa\label{f1}
\mathcal{F}_1&=&\frac1{24} \log \left( v_x^2 -w\, u_x^2\right)
\\
&=&-\frac1{12}\log x +\frac32 s^2 x + 189\, s^4 x^2 +\dots .
\nn
\eeqa
The genus two expression is more involved. Introduce two series
\beq\label{u1u2}
u_{1,2}=v\pm 2 \sqrt{w}=6\, s\, x + 324\, s^3 x^2 + 31104\, s^5 x^3 \pm
2\sqrt{x} \left( 1+18\, s^2 x +1458\, s^4 x^2+\dots\right).
\eeq
Then
\eqa
&&24^2\,\mathcal{F}_2=
\frac{{4\,u_1''}^3\,u_{12}}{5\,{u_1'}^4} -
  \frac{{4\,u_2''}^3\,u_{12}}{5\,{u_2'}^4}-
\frac{u_1''\,u_2''}{4\,u_1'\,u_2'}
\nn\\
&&\quad+
\frac{3\,u_1''}{4\,{u_1'}^3}\left(\frac12\,u_1''\,u_2'-\frac75\,u_1'''\,u_{12}
\right)
+
\frac{3\,u_2''}{4\,{u_2'}^3}\left(\frac12\,u_2''\,u_1'+\frac75\,u_2'''\,u_{12}
\right)\nn\\
&&\quad+
\frac1{4\,{u_1'}^2}\left(\frac{33}{10}\,{u_1''}^2-\frac9{10}\,u_1'''\,u_2'+
\frac1{10}\,u_1''\,u_2''+u_1^{IV}\,u_{12}\right)\nn\\
&&\quad+
\frac1{4\,{u_2'}^2}\left(\frac{33}{10}\,{u_2''}^2-\frac9{10}\,u_2'''\,u_1'+
\frac1{10}\,u_1''\,u_2''-u_2^{IV}\,u_{12}\right)\nn\\
&&\quad-
\frac1{4\,u_1'}\left(\frac{17}{5}\,u_1'''+\frac1{2}\,u_2'''\right)-
\frac1{4\,u_2'}\left(\frac{17}{5}\,u_2'''+\frac1{2}\,u_1'''\right)\nn\\
&&\quad-
\frac1{10\, u_{12}^2}\left(\frac{{u_1'}^3}{u_2'}+\frac{{u_2'}^3}{u_1'}
\right)
-\frac{1}{u_{12}^2}\left({u_1'}^2-\frac{11}{5}\,u_1'\,u_2'+{u_2'}^2\right)
\nn\\
&&\quad+
\frac{u_1''-u_2''}{u_{12}}\left(\frac{u_2'}{5\,u_1'}+\frac{u_1'}{5\,u_2'}+
1\right)
\\
&&
=-\frac1{240 x^2} +\frac{8505}2 s^6 x+\dots
\nn
\eeqa
Here $u_{12}=u_1-u_2$, $u_{1,2}'=\partial_x u_{1,2}$, $u_{1,2}''=\partial_x^2 u_{1,2}$ etc. Applying this algorithm one obtains the following table of the weighted numbers of triangulations of surfaces of genera 0, 1, 2 with $k\leq 20$ triangles ($k$ is necessarily even); the computation takes less than a second.

\begin{table}[!htbp]
\begin{center}\tiny 
    \begin{tabular} {|c|R{2.8cm}|R{3.0cm}|R{3.2cm}|N}
    \hline
    $k$ & $g=0$ & $g=1$ & $g=2$ & \\[8pt]
    \hline
    2 & 6 & 3/2 & 0  & \\[8pt]
    \hline
    4 & 216 & 189 & 0 & \\[8pt]
   \hline
    6 & 13608 & 26892 & 8505/2  & \\[8pt]
    \hline
    8 & 119744 & 4076568 & 2217618 &  \\[8pt]
    \hline
    10 & 540416448/5  & 3213210384/5  & 3905028468 &    \\[8pt]
    \hline
    12 & 11609505792 & 104047172352 & 231226436160 &  \\[8pt]
    \hline
    14 & 9425943686016/7 & 120228382104192/7 & 62004956093424 &  \\[8pt]
    \hline
    16 & 165505114570752 & 2877311706393600 & 15594280091334144 &  \\[8pt]
    \hline
    18 & 21285494650967040 & 487638320996544768 & 3749645355442763904 &  \\[8pt]
    \hline
    20 & 14195644503284514816/5 & 417102705028906942464/5 & 4360691488086816325632/5 &  \\[8pt]
    \hline
    \end{tabular}
\end{center}
\caption{Weighted triangle numbers $a_g\!\left(3^k\right)$.} 
\label{numbers-DZ}
\end{table}


In a similar way one can easily compute the weighted numbers of quadrangulations etc. of surfaces of genus 0, 1, 2. In principle one can extend this algorithm to higher genera but the calculations become more involved.

We plan to do large $g$ and large $k$ asymptotics of connected GUE correlators based on the two algorithms described above in an upcoming
 publication.

\appendix

\section{Appendix. GUE, Toda lattice and enumeration of ribbon graphs}\label{appa} \par

\setcounter{equation}{0}
\setcounter{theorem}{0}
\subsection{GUE partition function and orthogonal polynomials}

Consider the GUE partition function represented as an integral over the space ${\mathcal H}(N)$ of $N\times N$ Hermitean matrices $M=\left( M_{ij}\right)$
\beq\label{part}
Z_N({\bf s}; \epsilon)=\frac{(2\pi)^{-{N}} \epsilon^{-\frac1{12}}}{Vol(N)} \int_{{\mathcal H}(N)} e^{-\frac1{\epsilon} \tr \,V(M)} dM.
\eeq
Here the formal series $V$ depending on the parameters ${\bf s}=(s_3, s_4, \dots)$ has the form
\beq\label{pot}
V(M)=\frac12 M^2 -\sum_{j\geq 3} s_j  M^j.
\eeq
The integral with respect to the measure
$$
dM=\prod_{i=1}^N dM_{ii} \prod_{i<j} d\rea( M_{ij})\, d\ima (M_{ij})
$$
will be understood\footnote{An alternative way is to consider the integral \eqref{part} at $\epsilon=1$ as a formal expansion in the parameters $s_k$. Then one can extend $Z_N({\bf s}; 1)$ also to the variables $s_1$ and $s_2$, $V(M) =\frac12 M^2 -\sum_{j\geq 1} s_j M^j$. In this way one obtains a generating series for correlators of all traces $\tr\, M^j$ for arbitrary $j\geq 1$.} as a formal asymptotic expansion\footnote{Under certain assumptions for the polynomial $V(M)$ it can be rigorously justified \cite{EM, EMP} that the formal series considered below are asymptotic expansions of convergent matrix integrals.} with respect to the small parameter $\epsilon\to+0$. The pre-factor $Vol(N)^{-1}$ corresponds to the volume, with respect to the Haar measure, of the quotient of the unitary group over the maximal torus $\left[ U(1)\right]^N$
\beq\label{vol}
Vol(N)=Vol\left( U(N)/\left[ U(1)\right]^N\right)=\frac{\pi^{\frac{N(N-1)}2} }{G(N+1)}
\eeq
Here $G$ is the Barnes $G$-function taking the value
\beq\label{barnes}
G(N+1)=\prod_{n=1}^{N-1} n!
\eeq
at positive integers. The formula \eqref{vol} will be re-derived below.

Denote ${\mathcal D}_N$ the set of diagonal $N\times N$ matrices $\Lambda=\diag (\lambda_1, \lambda_2, \dots, \lambda_N)$ with real ordered eigenvalues $\lambda_1\leq \lambda_2\leq \dots \leq \lambda_N$. The map
\eqa\label{coord}
&&
U(N)/\left[ U(1)\right]^N\times {\mathcal D}_N\to {\mathcal H}(N)
\nn\\
&&
\\
&&
\left( U, \Lambda\right)\mapsto U\,\Lambda\, U^*
\nn
\eeqa
is a local diffeomorphism away from a subset of codimension three in ${\mathcal H}(N)$. Because of invariance of the measure w.r.t. to the action of unitary group one obtains
$$
\int_{{\mathcal H}(N)} e^{-\frac1{\epsilon} \tr \,V(M)} dM=Vol\left( U(N)/\left[ U(1)\right]^N\right)\int_{{\mathcal D}_N}  \Delta^2(\lambda) \, e^{-\frac1{\epsilon} \sum_{k=1}^N V(\lambda_k)}d\lambda_1\dots d\lambda_N.
$$
Here
$$
\Delta(\lambda)=\prod_{i<j} (\lambda_i-\lambda_j)
$$
is the Vandermonde determinant. Due to symmetry of the integrand one can rewrite the last formula as
$$
\int_{{\mathcal H}(N)} e^{-\frac1{\epsilon} \tr \,V(M)} dM=\frac1{N!}Vol\left( U(N)/\left[ U(1)\right]^N\right)\int_{\mathbb R^N}  \Delta^2(\lambda) \, e^{-\frac1{\epsilon} \sum_{k=1}^N V(\lambda_k)}d\lambda_1\dots d\lambda_N.
$$
Denote
\beq\label{ortopol1}
p_n(\lambda)=\lambda^n+a_{1n} \lambda^{n-1}+\dots+a_{nn}, \quad n=0, \, 1, \dots
\eeq
a system of monic polynomials orthogonal w.r.t. to the exponential weight
\beq\label{ortopol2}
\int_{-\infty}^\infty p_n(\lambda) \, p_m(\lambda) \, e^{-\frac1{\epsilon} V(\lambda)}d\lambda=h_n \, \delta_{mn}.
\eeq
Representing the Vandermonde as
$$
\Delta(\lambda)=\det  \left( \begin{array}{cccc} p_0(\lambda_1) & p_0(\lambda_2) & \dots & p_0(\lambda_N)\\
p_1(\lambda_1) & p_1(\lambda_2) & \dots & p_1(\lambda_N)\\
\cdot & \cdot & \dots & \cdot\\
\cdot & \cdot & \dots & \cdot\\
\cdot & \cdot & \dots & \cdot\\
p_{N-1}(\lambda_1) & p_{N-1}(\lambda_2) & \dots & p_{N-1}(\lambda_N)\end{array}\right)
$$
one obtains an expression of the last integral via the normalizing factors of the orthogonal polynomials
$$
\int_{\mathbb R^N}  \Delta^2(\lambda) e^{-\frac1{\epsilon} \sum_{k=1}^N V(\lambda_k)}d\lambda_1\dots d\lambda_N=N! \,h_0 \,  h_1\dots h_{N-1}.
$$
We conclude that
\beq\label{final}
\int_{{\mathcal H}(N)} e^{-\frac1{\epsilon} \tr \,V(M)} dM=Vol\left( U(N)/\left[ U(1)\right]^N\right) \, h_0 \, h_1\dots h_{N-1}.
\eeq
The formula \eqref{vol} for the volume $Vol\left( U(N)/\left[ U(1)\right]^N\right)$ can be easily derived from the last equation. Indeed, evaluating the lhs of eq. \eqref{final} at the Gaussian point ${\bf t}=0$ one obtains
$$
\int_{{\mathcal H}(N)} e^{-\frac1{2\epsilon} \tr\, M^2} dM=2^{\frac{N}2} \left( \pi\,\epsilon\right)^{\frac{N^2}2}.
$$
At ${\bf s}=0$ the orthogonal polynomials \eqref{ortopol1}--\eqref{ortopol2} are expressed via Hermite polynomials
$$
p_n(\lambda)=\epsilon^{\frac{n}2}\, {\rm He}_n(x), \quad \lambda=\epsilon^{\frac12}x.
$$
From
$$
\int_{-\infty}^\infty {\rm He}_n(x) \, {\rm He}_m(x) \, e^{-\frac12 x^2} dx=\sqrt{2\pi} \,n! \,\delta_{mn}
$$
it follows that
$$
h_n({\bf s}=0)= \epsilon^{n+\frac12} \sqrt{2\pi} \,n!
$$
So, eq. \eqref{final} at ${\bf s}=0$ takes the form
$$
2^{\frac{N}2} \left( \pi\,\epsilon\right)^{\frac{N^2}2}=Vol\left( U(N)/\left[ U(1)\right]^N\right)\cdot\left( 2\pi\right)^{\frac{N}2} \epsilon^{\frac{N^2}2} \prod_{n=1}^{N-1} n!
$$
This implies \eqref{vol}.

We conclude this Section with the following expression for the GUE partition function
\beq\label{gue}
Z_N({\bf s}; \epsilon) = h_0 \, h_1 \dots h_{N-1}.
\eeq
Our nearest goal is to prove that this partition function is the tau-function of a particular solution of Toda hierarchy.

\setcounter{equation}{0}
\setcounter{theorem}{0}
\subsection{GUE and Toda} \label{GUE-Toda-app}

Denote $v_n$, $w_n$ the coefficients of the three-term recursion relation for the orthogonal polynomials $p_n(\lambda)$
\beq\label{3term}
\lambda\, p_n(\lambda) =p_{n+1}(\lambda) +v_n \, p_n(\lambda) + w_n \, p_{n-1}(\lambda), \quad n\geq 0
\eeq
$p_{-1}=0$. That is, the orthogonal polynomials are eigenvectors of the second order difference operator
\beq\label{lax}
\left( L\, \psi\right)_n = \psi_{n+1} +v_n \, \psi_n + w_n \, \psi_{n-1}.
\eeq
The corresponding tri-diagonal matrix will also be denoted $L=\left( L_{ij}\right)$.

Denote
\beq\label{inner}
(f, g)=\int_{-\infty}^\infty f(\lambda) \, g(\lambda) \, e^{-\frac1{\epsilon} V(\lambda)}d\lambda
\eeq
an inner product on the space of polynomials. Recall that all integrals are understood as formal series in $\epsilon^{1/2}$ (actually, after division by $\sqrt{\epsilon}$ they contain only integer powers of $\epsilon$). The symmetry
$$
\left(\lambda\, p_n, p_m\right)=\left( p_n, \lambda\, p_m\right)~\Leftrightarrow~ L_{mn} h_m=L_{nm} h_n
$$
implies
\beq\label{wn}
w_n=\frac{h_n}{h_{n-1}}=\frac{Z_{n+1} Z_{n-1}}{Z_{n}^2}.
\eeq
Here $h_n=(p_n, p_n)$ (see eq. \eqref{ortopol2} above).

For an arbitrary square matrix $X=(X_{ij})$ denote $X_-$ and $X_+$ its upper- and lower-triangular parts
$$
X_-=\left( X_{ij}\right)_{i<j}, \quad X_+=\left( X_{ij}\right)_{i\geq j}, \quad X=X_+ + X_-.
$$

\begin{lemma} The orthogonal polynomials $p_n=p_n(\lambda)$ satisfy
\beq\label{dt}
\epsilon\,\frac{\pal p_n}{\pal s_j} +(A_j \, p)_n=0, \quad A_j = -\left( L^{j} \right)_-, \quad j\geq 3.
\eeq
\end{lemma}

\noindent \textit{Proof}. Write
$$
\frac{\pal p_n(\lambda)}{\pal s_j} =\sum_{i=0}^{n-1} A_{i\, n}^{(j)} \, p_i(\lambda), \quad n\geq 1
$$
for some coefficients $A_{in}^{(j)}$.
Differentiating in $s_j$ the equation $(p_n, p_m)=0$ for $m<n$ we obtain
$$
A_{mn}^{(j)}h_m +\frac1{\epsilon}  \left( \lambda^{j} p_n, p_m\right)=0.
$$
Introduce matrices of multiplication by powers of $\lambda$
\beq\label{multlam}
\lambda^{j}p_n(\lambda)=\sum_{i=0}^{n+j} \left(L^{j}\right)_{in} p_i(\lambda).
\eeq
We have
$$
\left( \lambda^{j}p_n, p_m\right)=\left( L^{j}\right)_{mn} h_m,
$$
hence
\beq\label{ab}
\epsilon\, A_{mn}^{(j)} =- \left( L^{j}\right)_{mn}, \quad m<n
\eeq
that is, 
$$
\epsilon\, A^{(j)}=- \left(L^{j}\right)_-.
$$ \epf

Repeating a similar calculation for $m=n$ we obtain
\beq\label{derh}
\frac{\pal}{\pal s_j} \log \frac{Z_{n+1}}{Z_n} \equiv \frac{\partial}{\partial s_j}\log h_n = \left( L^{j} \right)_{nn}.
\eeq

\begin{cor} \label{toda-gue-cor} The difference operator $L$ satisfies
\beq\label{todapol}
\epsilon\, \frac{\pal L}{\pal s_j} =\left[A_j,  L\right], \quad A_j=\left( L^{j}\right)_+.
\eeq
\end{cor}

\noindent \textit{Proof}. Differentiating equation
$$
\lambda\, p_n =\sum_{i=0}^{n+1} L_{in}  \,  p_i
$$
in $s_j$ and using  eq. \eqref{dt} obtain
$$
\epsilon\, \frac{\pal L}{\pal s_j} =\left[ L, \left( L^{j}\right)_-\right].
$$
Since the operators $L$ and $L^{j}$ commute we arrive at \eqref{todapol}. \epf

\begin{prop} \label{partition-tau-GUE-toda} 
The GUE partition function
$Z$ is a tau-function, in the sense of Definition \ref{defitau} of the Toda lattice hierarchy.
\end{prop} 
\noindent \textit{Proof}.
Cor.\,\ref{toda-gue-cor} tells that $w_n,v_n$ is a particular solution to the Toda lattice hierarchy.

It then follows from \eqref{wn} and \eqref{derh} that the partition function $Z_n$ satisfies 
eq.\,\eqref{taun2} and eq.\,\eqref{taun3}. Let $t_j=s_{j+1}, ~j=0,1,2,\dots.$ Here $s_1,s_2$ are 
understood as in the footnote 2.
Eq.\,\eqref{derh} implies that 
$$
\frac{\pal}{\pal t_{j}} \log \frac{Z_{n+1}}{Z_n}  = (j+1)\, h_{j-1}(n), \qquad j\geq 0
$$
where $h_{j-1}(n):=\frac{1}{j+1} (L^{j+1})_{nn}$. Define 
$$
\gamma_n(\lambda)= \frac1{\lambda} +  \sum_{j\geq 0} \frac{ (j+1)\, h_{j-1}(n-1)}{\lambda^{j+2}}.
$$
We have
$$
\sum_{j\geq 0}\frac{1}{\lambda^{j+2}} \frac{\pal}{\pal t_{j}} \log \frac{Z_{n+1}}{Z_n}  = \sum_{j\geq 0}\frac{1}{\lambda^{j+2}}  (j+1)\, h_{j-1}(n) = \gamma_{n+1}(\lambda)-\frac1{\lambda}.
$$
So
$$
\sum_{i,j\geq 0}\frac{1}{\mu^{i+2}\lambda^{j+2}} \frac{\pal^2}{\pal t_{j}\pal t_i} \log \frac{Z_{n+1}}{Z_n}  = \nabla(\mu)\, \gamma_{n+1}(\lambda).
$$
Here, $\nabla(\mu)$ is defined in \eqref{nabladef}. Noting that 
\eqa
 \nabla(\mu)\, \gamma_{n+1}(\lambda) & = & 
 \frac{\gamma_{n+1}(\lambda)(1+2\alpha_{n+1}(\mu))-\gamma_{n+1}(\mu)(1+2\alpha_{n+1}(\lambda))}{\lambda-\mu} +\gamma_{n+1}(\lambda)\gamma_{n+1}(\mu)\nn\\
 &=&  \frac{\tr \, R_{n+1}(\lambda) R_{n+1}(\mu) - \tr \, R_n(\lambda) R_n(\mu)}{(\lambda-\mu)^2}\nn
\eeqa
we obtain
$$
\sum_{i,j\geq 0}\frac{1}{\mu^{i+2}\lambda^{j+2}} \frac{\pal^2}{\pal t_{j}\pal t_i} \log Z_n = \frac{\tr \, R_n(\lambda) R_n(\mu) -1}{(\lambda-\mu)^2} .
$$
In the above formulae, $R_n$ is the matrix resolvent of $L$.  The proposition is proved.
\epf

\setcounter{equation}{0}
\setcounter{theorem}{0}
\subsection{GUE and enumeration of ribbon graphs}

After rescaling $M\mapsto \epsilon^{\frac12} M$ and expansion in powers of the parameters one obtains
$$
\int_{{\mathcal H}(N)} e^{-\frac1{\epsilon} \tr \,V(M)} dM=\epsilon^{\frac{N^2}2} \int_{{\mathcal H}(N)}  e^{-\frac12 \tr\, M^2 +\sum_{j\geq 3} \epsilon^{\frac{j}2-1} s_j\, \tr\, M^{j}}dM
$$
Expanding in a series in $\epsilon$ yields
\beq\label{pert}
\frac{Z_N({\bf s}; \epsilon)}{Z_N(0; \epsilon)}= \sum_{k\geq 0} \frac1{k!} \sum_{m\geq 0} \epsilon^m\sum_{i_1+\dots+i_k=k+2 m}
s_{i_1}\dots s_{i_k} \left\langle  \tr\, M^{i_1}\dots \tr\, M^{i_k} \right\rangle
\eeq
where
\beq\label{poln}
\left\langle  \tr\, M^{i_1}\dots \tr\, M^{i_k} \right\rangle:=\frac{\int  \tr\, M^{i_1}\dots \tr\, M^{i_k}\,e^{-\frac12 \tr\, M^2} \, dM}{\int  e^{-\frac12 \tr\, M^2}dM}.
\eeq
The coefficients \eqref{poln} of the perturbative expansion \eqref{pert} are polynomials in $N$ that can be computed by applying the Wick rule. E.g.,
$$
\left\langle \tr\, M^4\right\rangle=2N^3+N, \quad \left\langle \left(\tr\, M^3\right)^2\right\rangle=12 N^3 + 3 N, \quad \left\langle \tr\, M^6\right\rangle=5N^4 + 10 N^2
$$
etc.
Terms of the polynomial \eqref{poln} correspond to oriented ribbon graphs with $k$ vertices. 
Expansion of the logarithm of the partition function has a similar structure keeping connected graphs only:
\beq\label{logpert}
\log\frac{Z_N({\bf s}; \epsilon)}{Z_N(0; \epsilon)}= \sum_{k \geq 0} \frac1{k!} \sum_{m\geq 0} \epsilon^m\, \sum_{i_1+\dots+i_k=k+2 m}
s_{i_1}\dots s_{i_k} \left\langle  \tr\, M^{i_1}\dots \tr\, M^{i_k} \right\rangle_c.
\eeq

Introduce the 't Hooft coupling parameter
$$
x=N\, \epsilon.
$$
Re-expanding in $\epsilon$ the logarithm of the partition function we arrive at the main statement of this section, see \cite{BIZ}.

\begin{theorem} Logarithm of the tau-function of the solution to the Toda hierarchy given by the GUE partition function has the following expansion
\eqa\label{expan}
&&
\hspace{-4mm} \log Z_N({\bf s}; \epsilon)|_{N=\frac{x}{\epsilon}}=\log Z_{N}(0; \epsilon) +\sum_{g\geq 0} \epsilon^{2g-2} \mathcal{F}_g (x; s_3, s_4, \dots)
\nn\\
&&
\nn\\
&&
\hspace{-4mm}  \mathcal{F}_g(x; s_3, s_4, \dots )=\sum_{k\geq 0} \sum_{i_1, \dots, i_k} a_g(i_1, \dots, i_k) \, s_{i_1} \dots s_{i_k} x^h
\nn\\
&&
\nn\\
&&
\hspace{-4mm}  a_g(i_1, \dots, i_k) =\frac1{k!} \, 
\#\{\small{\mbox{connected oriented labelled ribbon graphs of genus } g \mbox{ with } k \mbox{ vertices of valencies } i_1, \dots, i_k}\}
\nn\\
&&\hspace{-4mm} \qquad\quad\quad ~~~~= \frac1{k!} \, \sum_{\Gamma}  \rho(\Gamma) = \sum_{\Gamma} \frac1{\#\, {\rm Sym} \, \Gamma}\\
&&\hspace{-4mm} h=2-2g -\left(k-\frac{|i|}2\right), |i|=i_1+\dots + i_k,\nn
\eeqa
where the two last summations are taken over all connected (unlabelled) ribbon graphs $\Gamma$ of genus $g$ with $k$ vertices of valencies $i_1$, \dots, $i_k$, $\rho(\Gamma)$ is the number of labelled ribbon graphs having the same topological shape $\Gamma$, and $\#\,{\rm Sym}\, \Gamma$ is the order of the symmetry group of $\Gamma$.
\end{theorem}

In the particular case $i_1=i_2=\dots=i_k=3$ the dual to the ribbon graph is a triangulation of the surface of genus $g$ consisting of $k$ triangles. Thus  $a_g\!\left( 3^k\right):=a_g(3, \dots, 3)$ ($k$ times) is equal to the weighted number of triangulations of genus $g$ with $k$ triangles. In a similar way, $a_g\!\left(4^k\right)$ is the weighted number of quadrangulations of a surface of genus $g$ with $k$ squares, etc.

Since
$$
Z_N(0; \epsilon)=(2\pi)^{-\frac{N}2} \epsilon^{\frac{N^2}2-\frac1{12}}\, G(N+1)
$$
we can also expand the first term in \eqref{expan} with the help of the asymptotic expansion of the Barnes $G$-function
$$
\log G(z+1)\sim \frac{z^2}2 \left( \log z -\frac32\right) +\frac{z}2\log{2\pi} -\frac1{12} \log z +\zeta'(-1) +\sum_{\ell\geq 1} \frac{B_{2\ell+2}}{4 \ell (\ell+1)z^{2\ell}}, \quad z\to\infty.
$$
This yields the following genus expansion of the logarithm of the tau-function of the interpolated Toda hierarchy where the shift operator $\psi_n\mapsto \psi_{n+1}$ acting on functions on a lattice is replaced with the translation $\psi(x) \mapsto \psi(x+\epsilon)$ acting on smooth functions on the real line,
\eqa\label{taugue}
\log\tau(x; s_3, s_4, \dots;\epsilon)&=&\frac{x^2}{2\epsilon^2} \left( \log x -\frac32\right) -\frac1{12} \log x +\zeta'(-1) +\sum_{g\geq 2} \epsilon^{2g-2} \frac{B_{2g}}{4g(g-1)x^{2g-2}}
\nn\\
&&
\nn\\
&&
+\sum_{g\geq 0} \epsilon^{2g-2} \mathcal{F}_g (x; s_3, s_4, \dots).
\eeqa

\begin{remark} The genus expansion \eqref{taugue} is often written as $1/N$ expansion, setting $x=1$, so $\epsilon=1/N$,
\eqa
&&
\log Z_N({\bf s}; N^{-1}) =-\frac{N^2}2 \log N-\frac{N}2\log 2\pi +\frac1{12}\log N+\log G(N+1) +\sum_{g\geq 0} N^{2-2g} {\cal F}_g (s_3, s_4, \dots)
\nn\\
&&
{\cal F}_g(s_3, s_4, \dots )=\sum_{k\geq 0} \sum_{i_1, \dots, i_k} a_g(i_1, \dots, i_k) \, s_{i_1} \dots s_{i_k}.
\nn
\eeqa
The coefficients $a_g(i_1, \dots, i_k)$ are the same as in \eqref{expan}.
\end{remark}

Observe that the coefficients of the connected correlators as polynomials in $N$ can be expressed via the numbers $a_g(i_1, \dots, i_k)$ enumerating ribbon graphs. Namely,
 $\forall\,k\geq 1$ and $\forall\, i_1,\dots,i_k$ such that $|i|$ is even, we have
$$\langle\tr \, M^{i_1} \, \tr \, M^{i_2} \, \dots \, \tr \, M^{i_k} \rangle_c\,=\,k!\,\sum_{0\leq g \leq \frac{|i|}4-\frac k2+\frac12} \, a_g(i_1,\dots,i_k) \, N^{2-2g-k+\frac{|i|} 2}.$$

\end{document}